\documentclass[%
 reprint,
 amsmath,amssymb,
 aps,
]{revtex4-2}

\usepackage{graphicx}
\usepackage{dcolumn}
\usepackage{bm}
\usepackage{subfigure} 

\begin{document}

\preprint{APS/123-QED}

\title{Core-softened water-alcohol mixtures: the solute-size effects}

\author{Murilo Sodré Marques}
\email{murilo.sodre@ufob.edu.br}
\affiliation{Centro das Ciências Exatas e das Tecnologias, Universidade Federal do Oeste da Bahia\\Rua Bertioga, 892, Morada Nobre, CEP 47810-059, Barreiras-BA, Brazil}



\author{Vinícius Fonseca Hernandes}
 
\affiliation{Departamento de Física, Instituto de Física e Matemática, Universidade Federal de Pelotas. Caixa Postal 354, 96001-970, Pelotas-RS, Brazil.}

\author{José Rafael Bordin}
\affiliation{Departamento de Física, Instituto de Física e Matemática, Universidade Federal de Pelotas. Caixa Postal 354, 96001-970, Pelotas-RS, Brazil.}
\date{\today}

\begin{abstract}
Water is the most anomalous material on Earth, with a long list of thermodynamic, dynamic and structural behaviors that deviate from what is expected. Recent studies indicate that the anomalies may be related to a competition between two liquids, which means that water has a potential liquid-liquid phase transition (LLPT) which ends at a liquid-liquid critical point (LLCP). In a recent work [\textit{J. Mol. Liq}, 2020, \textbf{320}, 114420], using molecular dynamics simulations and a core-softened potential approach, we have shown that adding a simple solute as methanol can "kill" the density anomalous behavior as the LLCP is suppressed by the spontaneous crystallization in a hexagonal closed packing (HCP) crystal near the LLPT. Now, we extend this work in order to realize how longer-chain alcohols will affect the complex behavior of water-alcohol mixtures in the supercooled regime. Besides core-softened (CS) methanol, ethanol and 1-propanol were added to a system of identical particles that interact through the continuous shouldered well (CSW) potential. We observed that the density anomaly gradually decreases its extension in phase diagrams until disappearing with the growth of the non-polar chain and the alcohol concentration, differently from the liquid-liquid phase transition (and the LLCP), which remained present in all analyzed mixtures, in according to \textit{Nature}, 2001, \textbf{409}, 692. For our model, the longer non-polar chains and higher concentrations gradually impact the competition between the scales in the CS potential, leading to a gradual disappearing of the anomalies until the TMD total disappearance is observed when the first coordination shell structure is also affected: the short-range ordering is favored, leading to less competition between short- and long-range ordering and, consequently, to the extinction of anomalies. Also, the non-polar chain size and concentration have an effect on the solid phases, favoring the hexagonal closed packed (HCP) solid and the amorphous solid phase over the body-centered cubic (BCC) crystal. These findings help elucidate the behavior of water solutions in the supercooled regime, and indicate that the LLCP can be observed in systems without density anomalous behavior.
\end{abstract}

\maketitle

\section{Introduction}
Life, as we know it, started and evolved in water solutions. We can say, therefore, that elucidate the demeanor of complex molecules in water, at both micro- and macro- levels, is of paramount importance in modern science \cite{rowlinson1982, franzese2008}. Although the behavior of complex biological proteins in water is a huge and complex problem, with many hydrophobic and hydrophilic sites, we can draw some information from simpler systems. In this way, a special class of aqueous solutions are those containing short-chain alcohols (i.e., alcohols with a small number of carbon atoms in the chain, like methanol, ethanol, or 1-propanol), most of which are miscible in water over the full range of concentrations \cite{koga2007,ruckenstein2009}. They have attracted a great attention of scientific community for decades for a number of reasons: (i) they are ubiquitous in the medical \cite{aspers2017}, food  \cite{nguyen2020}, transport \cite{paulina2019} and personal care industries \cite{vane2020}, among others; (ii) compared to water, the molecular structure of alcohols has an organic radical in place of one of the hydrogen atoms; as a consequence, alcohols do not form a fully developed hydrogen-bonded network, as in the case of water;  (iii) On the other hand, the presence of both the hydroxyl group and the organic radical, usually non-polar (amphiphilic character), allows interaction with a huge number of organic and inorganic compounds, making alcohols good solvents, since the solute-solvent interaction can have the same order of magnitude that the solute-solute and solvent-solvent interactions \cite{franks2000water}. In addition, (iv) although interactions with molecules of dual nature, such as alcohols, involve not only the hydrophobic hydration of the non-polar moiety of the molecule but also the hydrophilic interactions between the polar groups and water molecules, they still constitute a model for the investigation of the hydrophobic effect \cite{noel2002,Xi2016}.

With water as a solvent, both the complexity of the analysis and the richness of phenomena observed for such solutions are highlighted. Water is the most anomalous material, with more than 70 known anomalies~\cite{url}. Probably, the most well-known is the density anomaly. While most materials increases density upon cooling, liquid water density decreases when cooled from 4$^o$C to 0$^o$C at atmospheric pressure~\cite{Ke75}. Recent findings have indicated a relation between the anomalies and another unique feature of water: the liquid polymorphism and the phase transition between two distinct liquids~\cite{gallo2016}. In addition to the usual liquid-gas critical point (whose near-critical properties are so drastically different from those of liquid water \cite{anisimov2004}), since the 1990s \cite{poole1992} the existence of a second critical point - the liquid-liquid critical point (LLCP)  has been hypothesized by simulations and, since then, it has been subject of extensive debate~\cite{Limmer11, kesselring2012, Limmer13, poole13,Palmer13a, Palmer18}. It has not yet been reported from experiments since it is located in the so-called no-man's land: due to the spontaneous crystallization it is (almost) impossible to reach this region through experiments - however, some recent experiments show strong evidence of the existence of the LLCP~\cite{Caupin15, Taschin13, Hestand18} , as the works of Kim et al~\cite{kim2017, kim2020}. The LLCP is located at the end of a first order liquid-liquid phase transition (LLPT) line between low-density liquid (LDL) and high-density liquid (HDL) at low temperatures \cite{poole1992, stanley1997, stanley1998, stanley2000, sciortino2003, debenedetti2003, stanley2003a, Stanley2005, gallo2016, handle2017}. Unlike the liquid–gas phase transition (LGPT), which always has a positive slope of the first-order phase transition line because $\Delta S>0$ and $\Delta V>0$ (Clausius–Clapeyron relation), the LLPT line can indeed be either negative or positive, depending on the atomic pair interactions of the model \cite{gibson2006,franzese2011}. 

In this regard, the critical behavior of mixtures has been widely analyzed from both an experimental and theoretical point of view \cite{forman1962, stilinger64, sengers1983, chang1986, ludmer1987, prausnitz2000, javier2006, artemenko2010, yamamoto2011, bell2017}. Studies on the effects of solute size on the thermodynamic and structural properties of the aqueous solutions have been done \cite{harris1998, ashbaugh2001, chowdhuri2006, okamoto2018,gao2020} and, according to the authors' knowledge, there are still no analysis of alcohol chain  size influence on the critical properties and on polymorphism of these mixtures. In this way, some questions that arise are: how longer alcohol chains will affect the behavior of a system of identical particles that interact through the continuous shouldered well (CSW) potential in the supercooled regime? How possible changes in the critical behavior can be related to the anomalous behavior? Also, there is some effect in the spontaneous crystallization observed near the LLCP?

To answer these questions, we perform extensive simulations in order to analyze the phase diagram of water and water-alcohols mixtures, where both water and hydroxyls are modeled by core-softened potentials, which already have a long history of describing water anomalies both in bulk and confined environments\cite{jagla1998, jagla1999, alan2008a,Meyer99,krott15, bordin2018}, despite the fact that such isotropic potentials are not water because of their lacking of hydrogen-bond directionality \cite{franzese2011}. In our very recent work~\cite{marques2020} we have explored the supercooled regime of pure water, pure methanol and their mixtures: hydroxyl groups of different molecules interact through CSW potential~\cite{franzese2007, franzese2008} (which has already been studied, also analytically, and the occurrence of a third critical point was predicted \cite{artemenko2008}), while all other interactions are reproduced by LJ-like contributions, and our main focus in that publication was on the relations between density anomaly, liquid-liquid phase transition and spontaneous crystallization by means of Molecular Dynamics simulations. In the present work, in a similar way to that performed by Urbic et al\cite{urbic2015jcp}, we extend this scheme to ethanol ($CH_3-CH_2-OH$) and 1-propanol ($CH_3-CH_2-CH_2-OH$), by modeling such alcohols as linear chains constituted by three (trimers) and four (tetramers) partially fused spheres, respectively, and the primary objective is to complement our previous study by exploring the influence of the size of the solute on the thermodynamic properties of mixtures of core-softened water and alcohols: methanol, ethanol and 1-propanol. 'CSW' by chance coincides with the acronym of Core-Softened Water, but the reader must keep in mind the original meaning of CSW and its difference with water, whose entropy decreases when the molecular volume increases, while in CSW potential the entropy behaves like in usual liquids (argon-like) \cite{franzese2011}.  

The remaining of the paper is organized as follows. In Section II we present our interaction models for water and alcohols molecules, and summarize the details of the simulations. Next, in section III our most significant results for our CS mixture  model  are introduced. In particular, we will focus on the concentration dependence of the LLCP and influence on the excess entropy of mixture in comparison with pure CSW potential. The paper is closed with a  brief summary of our main conclusions and perspectives.

\section{Model and simulation details}

\begin{figure}[ht!]
\qquad
\centering
\subfigure[]{\includegraphics[width=0.2\textwidth, height=0.15\textwidth]{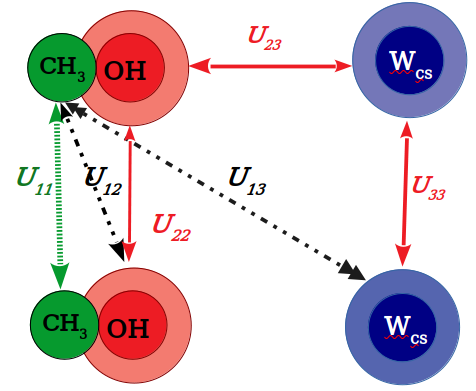}}
\qquad
\centering
\subfigure[]{\includegraphics[width=0.2\textwidth, height=0.15\textwidth]{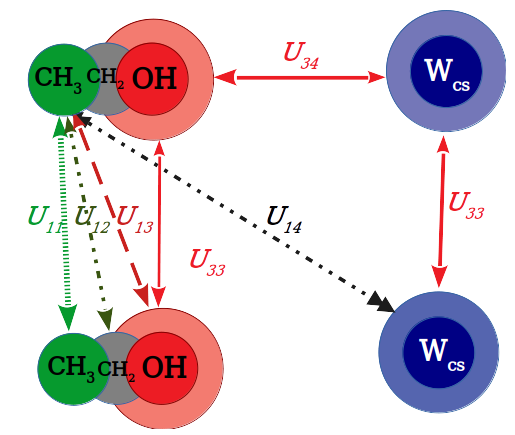}}
\qquad
\centering
\subfigure[]{\includegraphics[width=0.2\textwidth, height=0.15\textwidth]{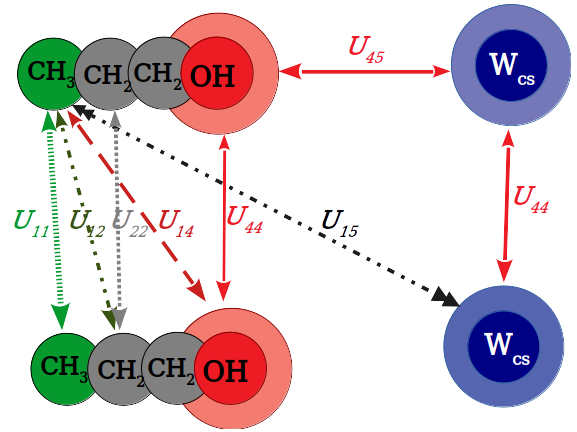}}
\qquad
\centering
\subfigure[]{\includegraphics[width=0.2\textwidth, height=0.15\textwidth]{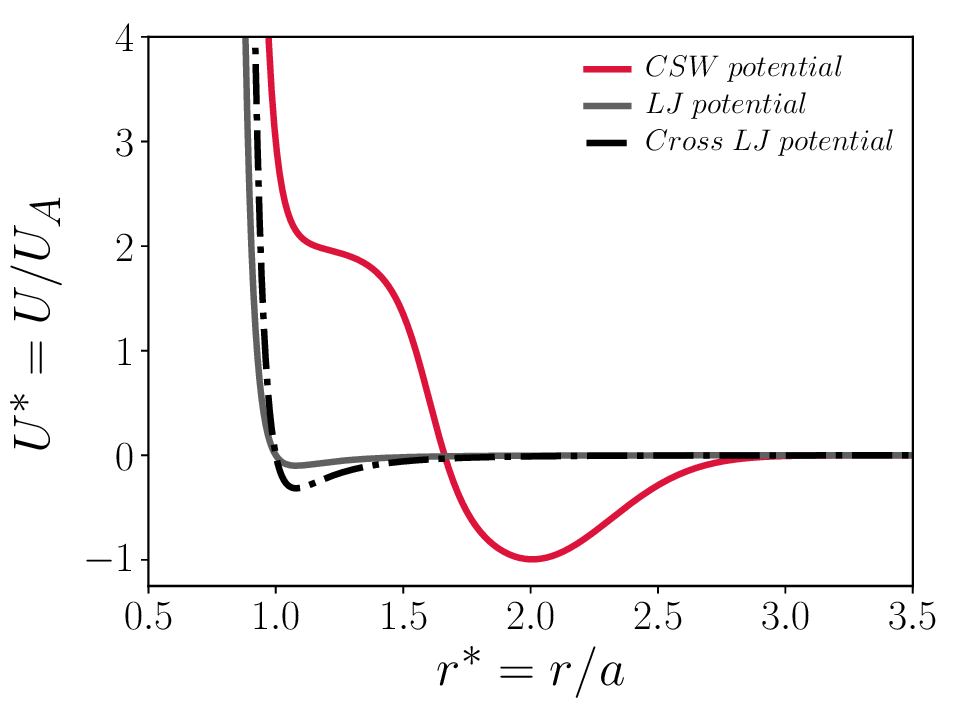}}
\caption{In (a), (b) and (c), our model for methanol, ethanol and 1-propanol is outlined, while in (d) we see the interaction between water and hydroxyl's is described by the CSW potential, while other interactions behave like a 24-6 Lennard-Jones potential.}
\label{fig1}
\end{figure}

Our water-like solvent will be  modeled by a 1-site  core-softened fluid in which particles interact with the potential model proposed by Franzese~\cite{franzese2007}.  Water-like particles $W_{CS}$ are represented by spheres with a hard-core of diameter $a$ and a soft-shell with radius $2a$, whose interaction potential is given by 
\begin{eqnarray}
U^{CS}(r) & = &\frac{U_R}{1+exp\left[\Delta(r-R_{R})\right ]}\nonumber \\  
& & -  U_{A}exp\left ( -\frac{(r-R_A)^2}{2\delta_A^2} \right )+U_A\left ( \frac{a}{r} \right )^{24}.
\label{franzese}
\end{eqnarray}
\noindent With the parameters $U_R/U_A=2$, $R_R/a=1.6$, $R_A/a=2$, $(\delta_A/a)^2=0.1$, and $\Delta=15$ this potential displays an attractive well for $r\sim 2a$ and a repulsive shoulder at $r\sim a$, as can be seen  in figure~\ref{fig1}(d) (red curve). All quantities with an asterisk are in reduced dimensionless units. The competition between these two length scales leads to water-like anomalies, as the density and the diffusion anomalies, and to the existence of a liquid liquid critical point~\cite{ franzese2007,alan2008,franzese2010,franzese2011}.

As a direct extension and building on our previous work on mixtures of water and methanol, where a single methanol molecule was constituted by two tangent spheres, in this work we have followed the extension made by Urbic {\it et al.} towards ethanol ($CH_3-CH_2-OH$) and 1-propanol ($CH_3-CH_2-CH_2-OH$) being modeled as linear trimers and tetramers, respectively (Fig. \ref{fig1}). Trimers are constructed as linear rigid molecules consisting of three partially fused spheres, where two adjacent spheres are placed at fixed distance $L_{ij} = 0.60a$, with $a$ as the unit length. Analogously, tetramers are modeled as linear rigid molecules consisting of four partially fused spheres, where two adjacent spheres are placed at fixed distance $L_{ij} = 0.78a$. In all models, hydroxyl groups interact through a soft-core potential (eq \ref{franzese}), while $CH_2$ and $CH_3$ groups are nonpolar and interact through LJ-like potential,

\begin{equation}
U^{LJ}=\frac{4}{3}2^{2/3}\epsilon \left [ \left ( \frac{\sigma}{r} \right )^{24} -\left ( \frac{\sigma}{r} \right )^{6} \right ],
\label{LJ246}
\end{equation}
\noindent $OH-CH_2$ and $OH-CH_3$ interactions are of $LJ$ type as well. $LJ$ parameters are reported in Table \ref{table1}. As in last work, quantities are reported in reduced dimensionless units relative to the hydroxyl group diameter and the depth of its attractive well.

\begin{table}[ht!]
    \centering
    \begin{tabular}{|c|c|c|c|c|c|}
    \hline
 & $L_{ij}$  & $\epsilon_{1n}$  & $\epsilon_{nn}$  & $ \sigma_{1n}$ & $ \sigma_{nn}$ \\ \hline
 Dimers & 1.000 & $0.316$  & $0.100$ & $1.000$  & $1.000$ \\
 Trimers & 0.600 & $0.400$  & $0.400$ & $1.115$  & $1.230$ \\
 Tetramers & 0.780 & $0.500$  & $0.500$ & $1.115$  & $1.230$ \\ \hline
 
    \end{tabular}
    \caption{Potential parameters for dimers ($n = 2$), trimers ($n = 2, 3$), and tetramers ($n = 2, 3, 4$); the $OH$ group is labeled with $1$. As for the bond
length, $j = i + 1$ \cite{urbic2015jcp}.}
    \label{table1}
\end{table}

\setcounter{subfigure}{0}

The simulations were performed in the $NPT$ ensemble with a fixed number  of molecules ($N_{tot} = 1000$). $N_{alc} = \chi_{alc} N_{tot}$ is the number of alcohol (methanol, ethanol or propanol) molecules and $N_w = N_{tot}-N_{alc}$ that of water molecules, where $\chi_{alc}$ is the alcohol mole fraction, which has been varied from 0.0 (pure water), 0.01, 0.05 and 0.1 (we've focused in low-concentration range). The temperature and pressure were controlled using the optimized constant pressure stochastic dynamics proposed by Kolb and D\"unweg~\cite{Kolb99} as implemented in the ESPResSo package~\cite{arnold2006,arnold2013}. This barostat implementation allows for the use of a large time step. This was set to $\delta t^* =0.01$, and the equations of motion were integrated using the velocity Verlet algorithm. The Langevin thermostat~\cite{allen2017}, that keeps the temperature fixed, has a coupling parameter $\gamma_0=1.0$. The piston parameters for the barostat are $\gamma_p=0.0002$ and mass $m_p = 0.001$. Here, the molecule density $\rho$ is defined as $N_m/<V_m>$ with $<V_m>$ being the mean volume at a given pressure and temperature. The pressure was varied from $P^* = 0.005$ up to $P^* = 0.30$ with distinct intervals - a finer grid was used in the vicinity of the critical points. The isobars were evaluated by cooling the system from $T^*= 0.70$ up to $T^* = 0.20$. A random system is created at the highest temperature along the isobar - at this temperature the system is always in fluid state, and then we run the dynamics for $5 \times 10^6$ time steps in the $NVT$ ensemble to thermalize the system. This was  followed by $1 \times 10^6$ time steps in the $NPT$ ensemble to equilibrate the system's pressure and density and $1 \times 10^7$ time steps further for the production of the results, with averages and snapshots being taken at every $1 \times 10^5$ steps. After that, we cool the system to the next lower temperature $T - \delta T$ in the isobar, repeat the thermalization and the equilibration process in the $NVT$ and $NPT$ ensemble, so the system can relax at the new temperature and achieve the proper new density, and then run the production steps. In this way, the two equilibration steps allow the system to cool down and relax at the new temperature close the previous one. To ensure that the system is in the equilibrium we averaged quantities as energy, entalpy, density, instantaneous pressure and temperature along the simulations, including in the equilibration processes.

Also in a similar way to the previous work, we evaluated the temperature of maximum density (TMD) and the locus of the maximum of response functions close to the critical point at the fluid phase (the isothermal compressibility $\kappa_T$, the isobaric expansion coefficient $\alpha_P$ and the specific heat at constant pressure $C_P$):

\begin{equation}
	\begin{split}
    \kappa_T & = \frac{1}{\rho} \left ( \frac{\partial \rho}{\partial P} \right )_T, \\
     \alpha_P & = -\frac{1}{\rho} \left (\frac{\partial \rho}{\partial T} \right )_P, \\
      C_P & = \frac{1}{N_{tot}} \left(\frac{\partial H}{\partial T} \right)_P,
    \end{split}
\end{equation}

\noindent where $H = U + PV$ is the system enthalpy, with $V$ the mean volume obtained from the $NPT$ simulations. The quantities shown in the Electronic Supplementary Material (ESI)$^\dag$ were obtained by numerical differentiation. As consistency check, we have obtained the same maxima locations when using statistical fluctuations: the compressibility is a measure of volume fluctuations, the isobaric heat capacity is proportional to the entropy fluctuations experienced by N molecules at fixed pressure, and the thermal expansion coefficient reflects the correlations between entropy and volume fluctuations ~\cite{allen2017,tuckerman2010}. 

In order to describe the connection between structure and thermodynamics, we have analyzed the  radial distribution function (RDF) $g(r^*)$, which was subsequently used to compute the excess entropy. $s_{ex}$ can be obtained by counting all accessible configurations for a real fluid and comparing with the ideal gas entropy \cite{dzugutov1996}. Consequently, the excess entropy is a negative quantity since the liquid is more ordered than the ideal gas. Note  that $s_{ex}$ increases  with  temperature  just  like  the  full entropy $S$ does; in fact $s_{ex} \rightarrow 0$ as temperature goes to infinity at fixed density because the system approaches an ideal gas \cite{dyre2018,dyre2020}. Analytically, the excess entropy may be computed if the equation of state is known \cite{galliero2011}. A  systematic expansion of $s_{ex}$ exists in terms of two-particle, three-particle contributions, etc.,

\begin{equation}
	s_{ex} = s_2 + s_3 + s_4+... 
\end{equation}

The  two-particle  contribution  is  calculated  from  the  radial distribution function $g(r)$ as follows:

\begin{equation}
s_2=-2\pi \rho \int_{0}^{\infty}\left [ g(r)\ln g(r) - g(r)+1) \right ]r^2dr,
\end{equation} 

since $s_2$ is the dominant contribution to excess entropy \cite{raveche1971,baranyai1989} and it is proved to be between $85\%$ and $95\%$ of the total excess entropy in Lennard-Jones systems \cite{sharma2006}. Also, the translational order parameter $\tau$ was evaluates. It is defined as~\cite{Er01}
\begin{equation}
\label{order_parameter}
\tau \equiv \int^{\xi_c}_0  \mid g(\xi)-1  \mid d\xi,
\end{equation}
\noindent where $\xi = r\rho^{1/3}$ is the interparticle distance $r$ scaled with the average separation between pairs of particles $\rho^{1/3}$. $\xi_c$ is a cutoff distance, defined as $\xi_c = L\rho^{1/3}/3$, where $L$ is the simulation box size. For an ideal gas (completely uncorrelated fluid), $g(\xi) = 1$ and $\tau$ vanishes. For crystals or fluids with long range correlations $g(\xi) \neq 1$ over long distances, which leads to $\tau >0$. The excess entropy and the translational order parameter $\tau$ are linked for the reason that both are dependent on the deviation of $g(r)$ from unity. 

Another structural quantity evaluated was the orientational order parameter (OOP), that gives insight on the local order~\cite{steinhardt1983, Er01, DeOliveira2006, Yan05}. The OOP for a specific particle $i$ with $N_b$ neighbors, is given by
\begin{equation}
\label{OOP_ql}
q_{l} (i) = \sqrt{\frac{4 \pi}{2l + 1} \sum_{m = -l}^{l} \vert q_{lm} \vert^2 },
\end{equation}
with
\begin{equation}
\label{OOP_qlm}
q_{lm} (i) = \sqrt{\frac{1}{N_{b}} \sum_{j=1}^{N_{b}} Y_{lm}(\theta(\vec{r}_{ij}),\phi(\vec{r}_{ij})) }.
\end{equation}
where $Y_{lm}$ are the spherical harmonics of order $l$ and $\vec{r}_{ij}$ is the distance between two CSW-particles, regardless in which molecule they are. The OOP for a whole system is obtained taking the average over the parameter value for each particle $i$, $q_l = \langle q_l (i) \rangle _i $. In this work we evaluated the OOP for $l = 6$, using the freud python library \cite{freud2020}, and the number of neighbors for each particle was found computing Voronoi diagrams using voro++ \cite{voro++}.

The dynamic behaviour was analyzed by the mean square displacement (MSD), given by
\begin{equation}
\label{r2}
\langle [\vec r(t) - \vec r(t_0)]^2 \rangle =\langle 
\Delta \vec r(t)^2 \rangle\;,
\end{equation}
where $\vec r(t_0) =$ and  $\vec r(t)$ denote the particle position at a time $t_0$ and at a later time $t$, respectively. The MSD is then related to the diffusion coefficient $D$ by the Einstein relation,
\begin{equation}
 D = \lim_{t \rightarrow \infty} \frac{\langle \Delta 
\vec r(t)^2 \rangle}{6t}\;.
\end{equation}
\noindent For alcohol molecules we have considered the center of mass displacement. Then, the diffusion anomaly region was obtained via the $(D^*, P^*)$ isotherms. For regular fluids, it is expected that $D^*$ decreases with $P^*$. However, anomalous fluids (as water and the CSW fluid) have diffusion anomaly, characterized by an increase in $D^*$ with $P^*$. With this, the diffusion anomaly extrema (DE) may be defined using the minima and maxima in these curves~\cite{alan2008,franzeseJCP2010,marques2020a}.

The onset of crystallization was monitored analyzing the local structural environment of particles  by means of the Polyhedral Template Matching (PTM) method implemented in the Ovito software~\cite{Larsen2016,ovito}. Ovito was also employed to visualize the phases and take the system snapshots.


\section{Results and discussion}

\subsection{Pure CSW potential phase diagram}

\begin{figure}[h!]
    \centering
     \subfigure[]{\includegraphics[width=0.39\textwidth]{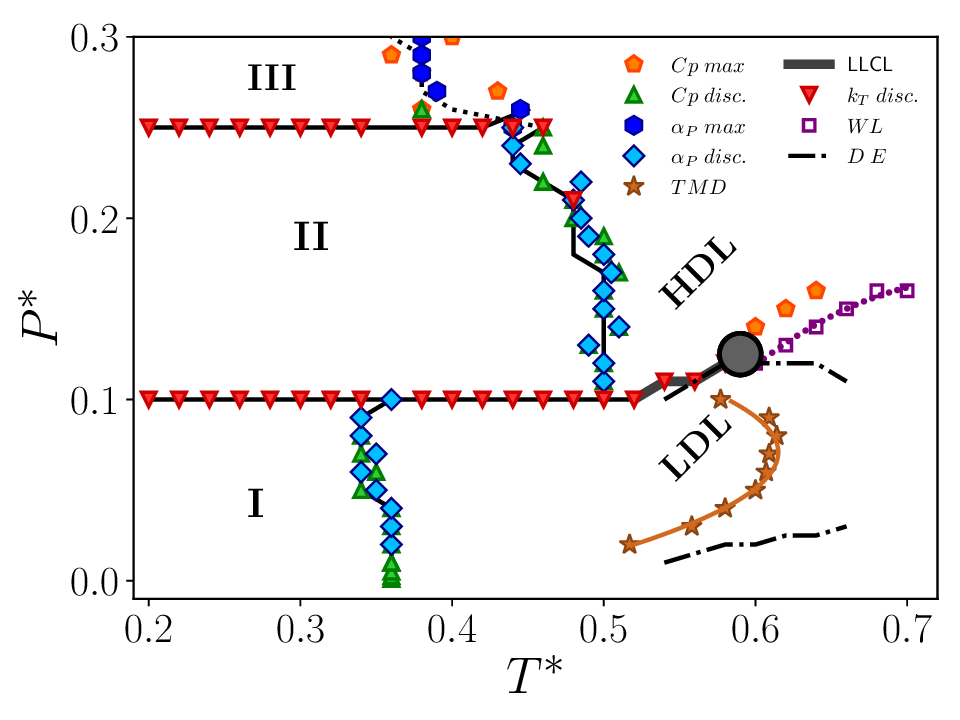}}
    \centering
     \subfigure[]{\includegraphics[width=0.225\textwidth]{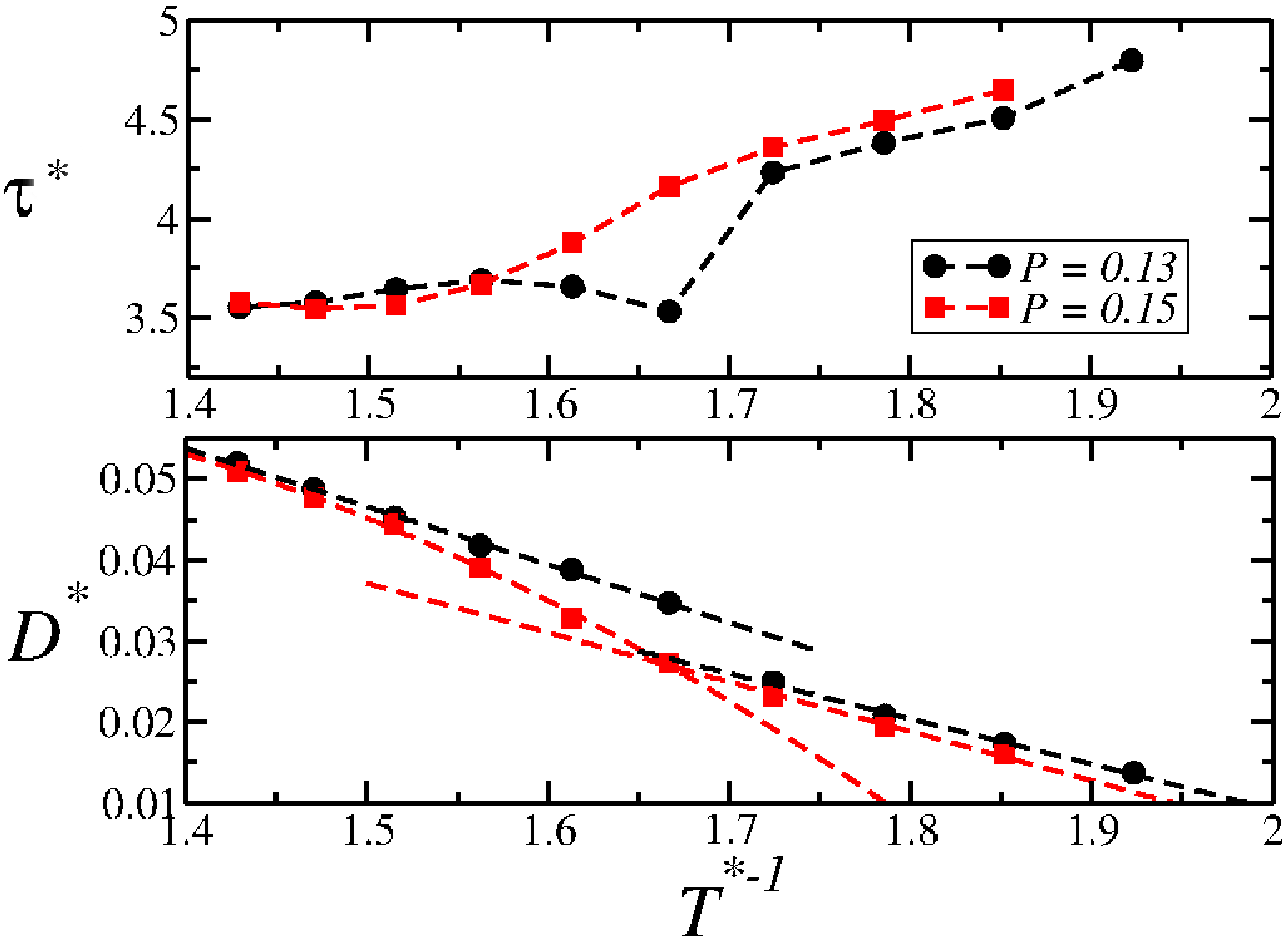}}
    \centering
     \subfigure[]{\includegraphics[width=0.24\textwidth]{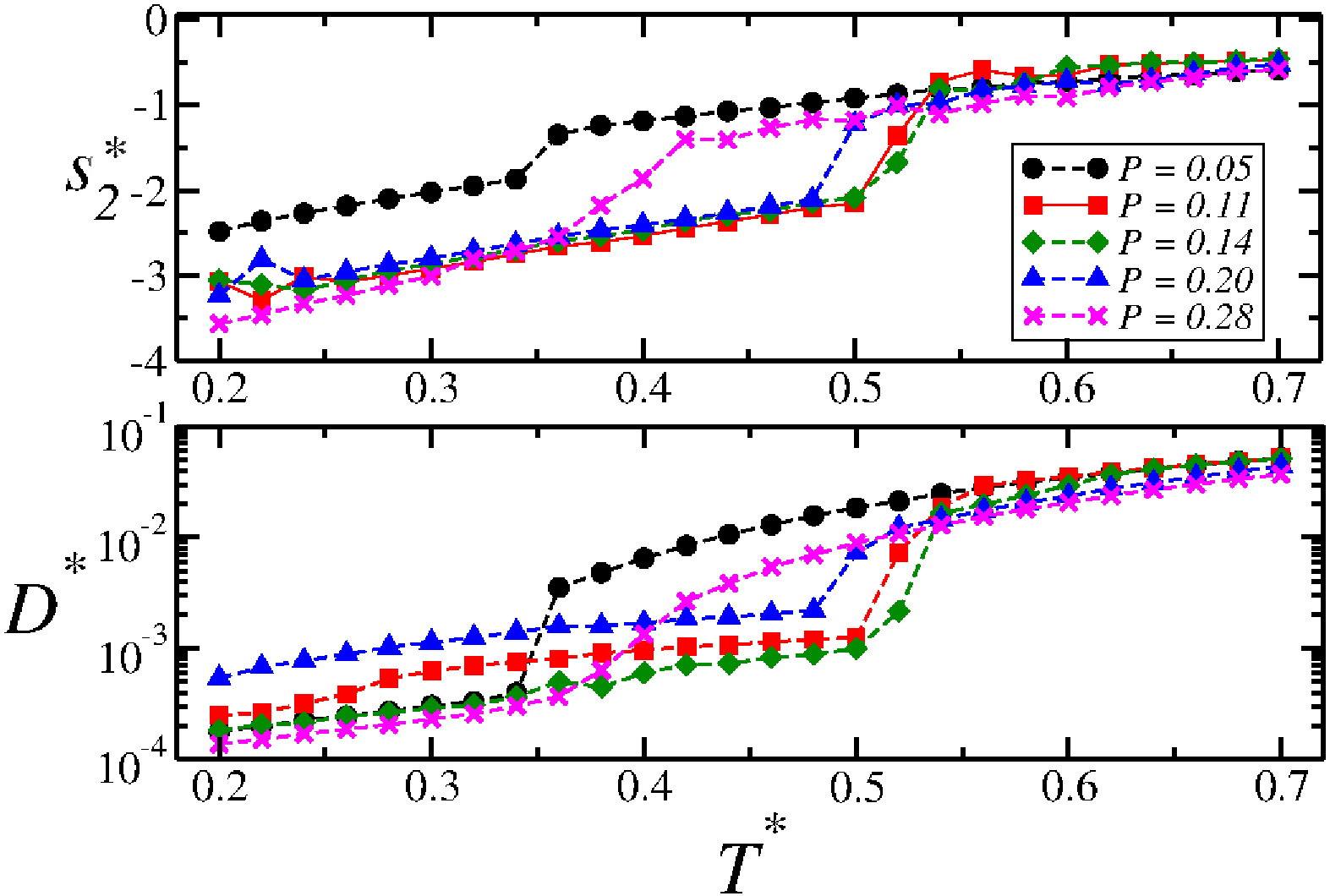}}
    
    \caption{(a) $PT$ phase diagrams for a system of identical particles that interact through the CSW potential showing the solid phases I (BCC solid), II (HCP solid) and III (amorphous solid) and the low and high density liquid phases. The points in the phase separations indicate distinct discontinuities or maxima in the evaluated response functions. The Widom Line (WL) is the extension of the LLPT curve into the one-phase region and the locus of maximum fluctuations of the order parameter \cite{holten2012, bianco2019}. The TMD and DE lines delimit the region of density and diffusion anomaly, respectively. (b) the translational order parameter $\tau$,the pair excess entropy (not shown here for simplicity) and the diffusion coefficient $D^*$ have discontinuities in the LDL-HDL transition for the subcritical isobar $P^*=0.13$, and a fragile to strong transition for the supercritical isobar $P^*=0.15$ as it crosses the Widom Line.
(c) The solid-liquid coexistence lines were draw based in the discontinuities in the pair excess entropy, the translational order parameter (not shown here for simplicity) and in the diffusion constant.}
    \label{waterPTdiagram}
\end{figure}

\begin{figure}[h!]
    \centering
     \subfigure[]{\includegraphics[width=0.4\textwidth]{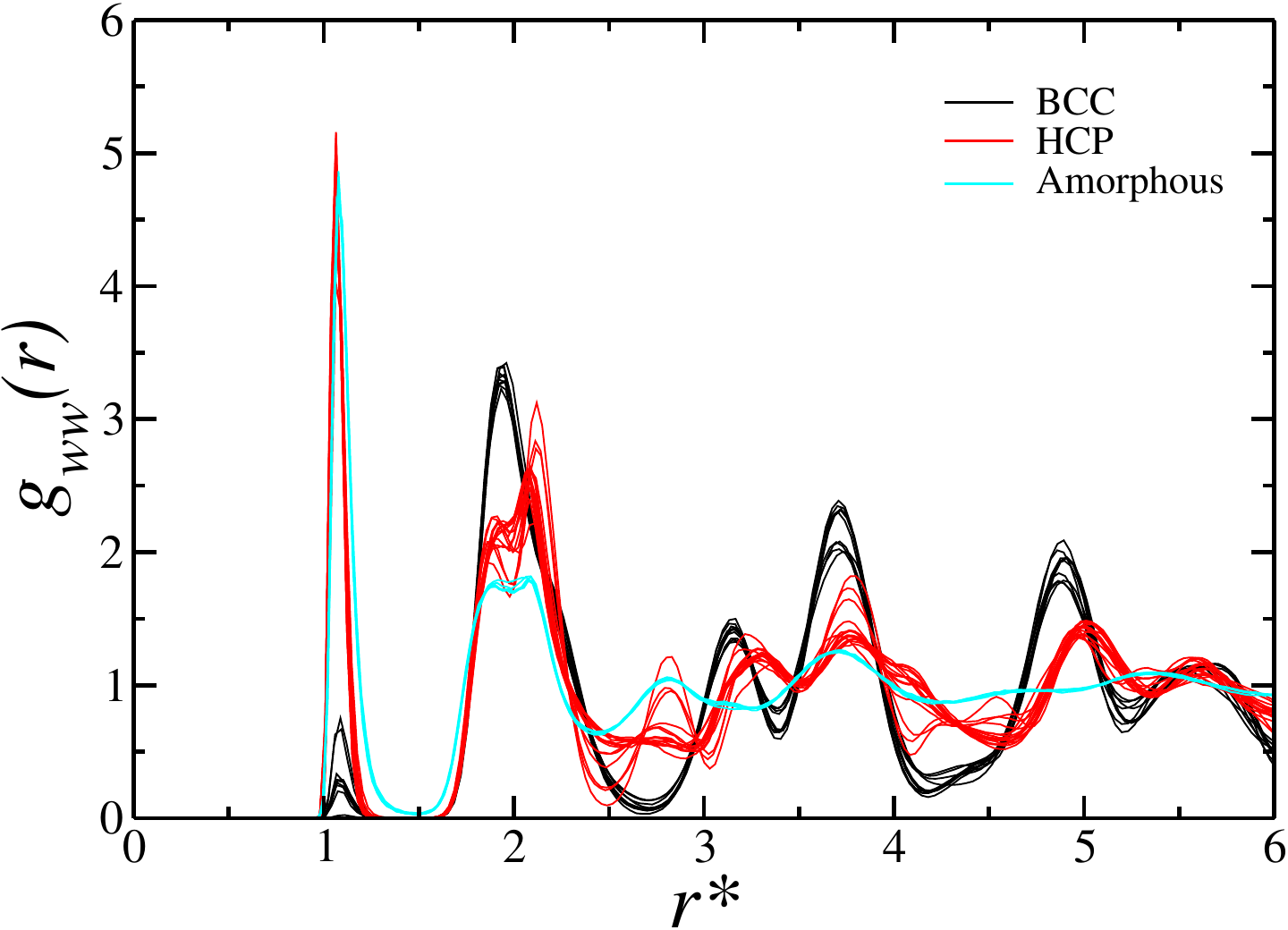}}
    \centering
     \subfigure[]{\includegraphics[width=0.115\textwidth]{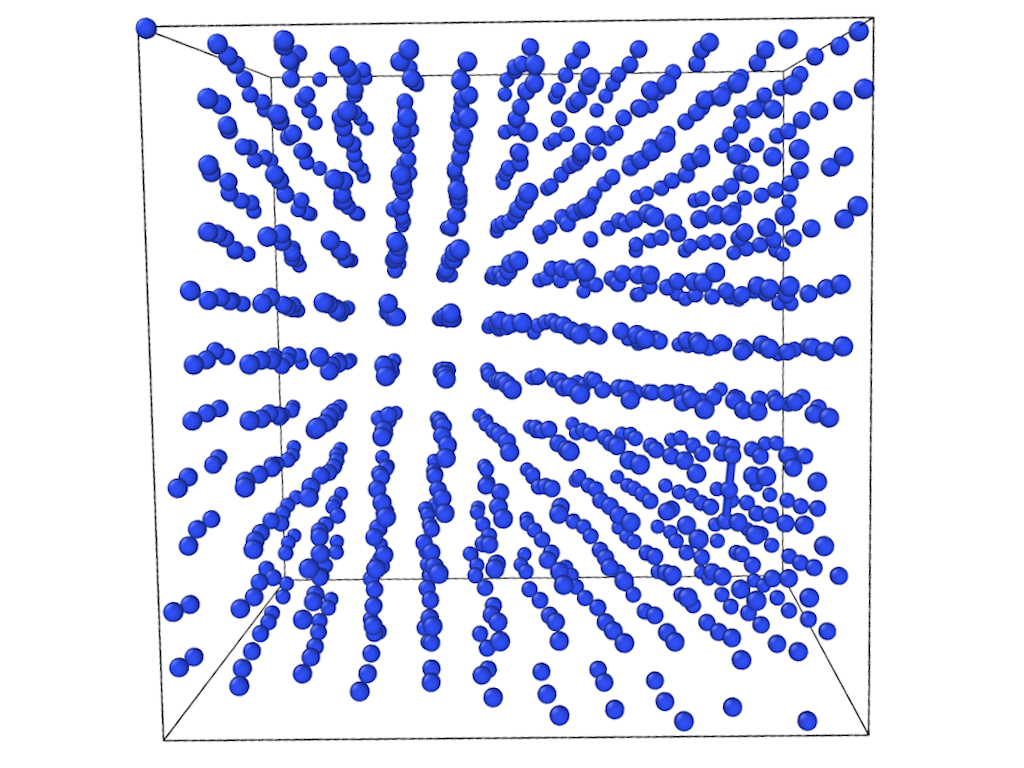}}
    \centering
     \subfigure[]{\includegraphics[width=0.115\textwidth]{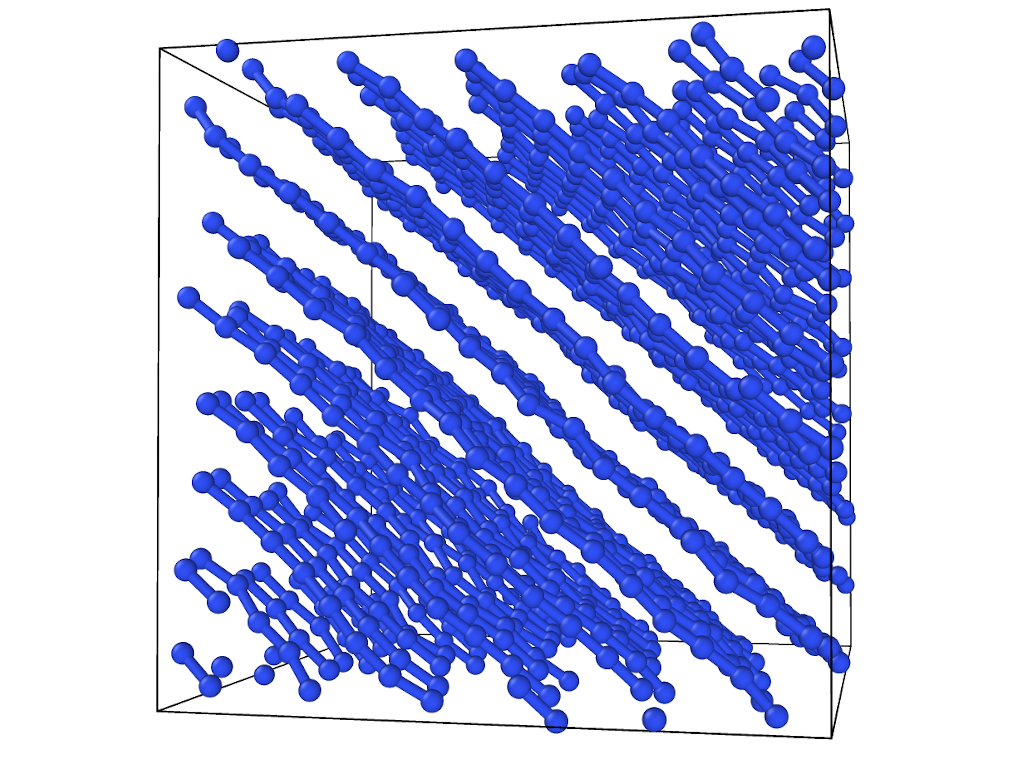}}
    \centering
     \subfigure[]{\includegraphics[width=0.115\textwidth]{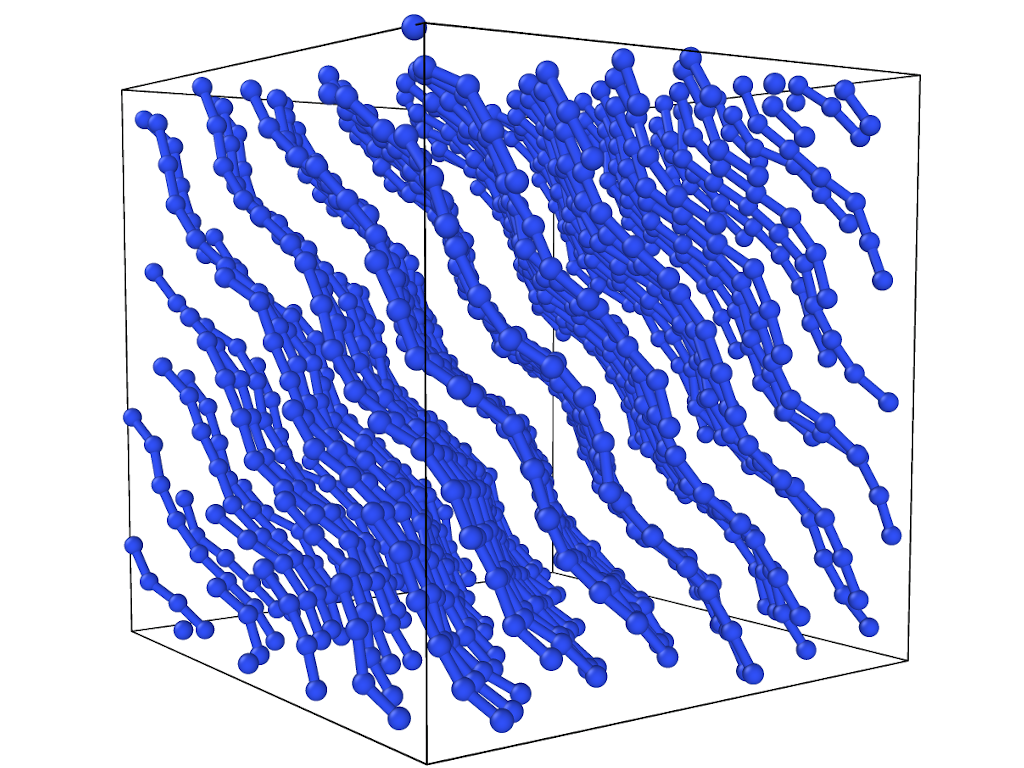}}
     \centering
     \subfigure[]{\includegraphics[width=0.115\textwidth]{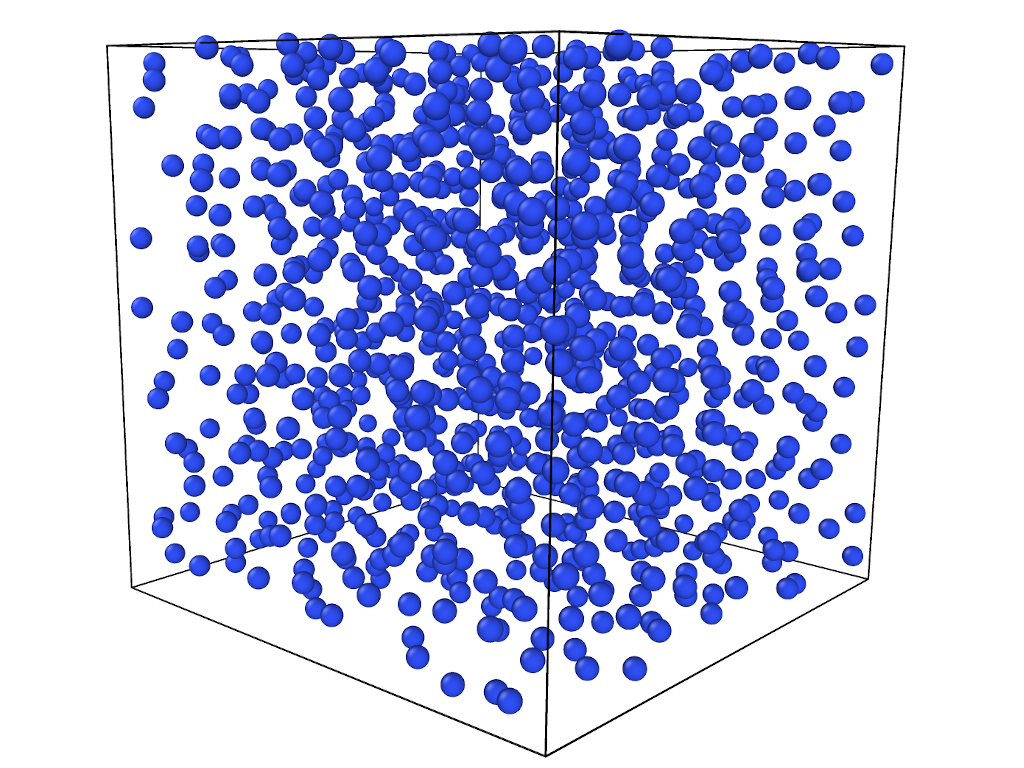}}
     \subfigure[]{\includegraphics[width=0.4\textwidth]{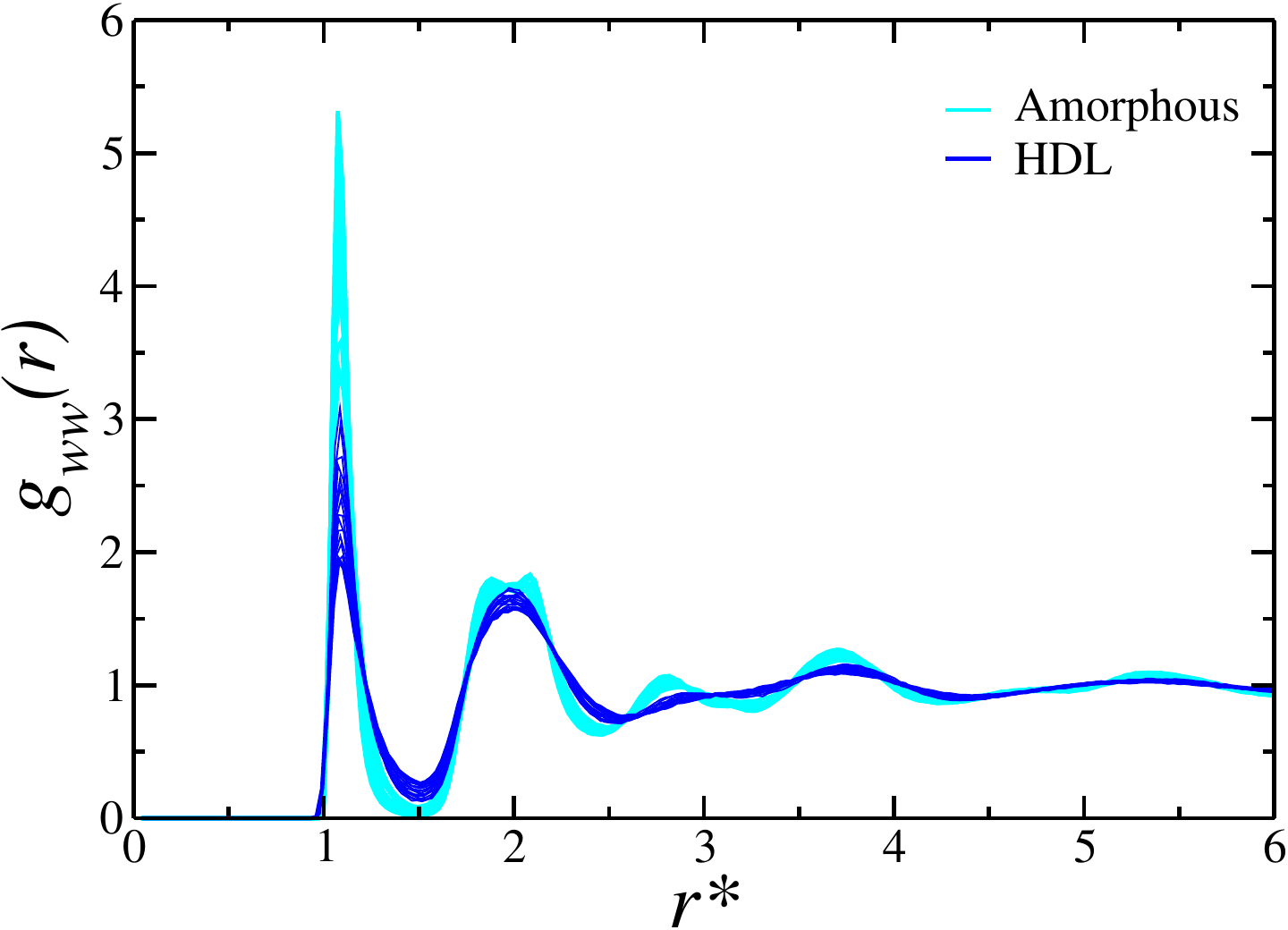}}
    \caption{(a) Pure CSW potential radial distribution function (RDF) along the isotherm $T^* = 0.26$. Black lines are the pressures in the BCC phase, red lines in the HCP phase and cyan in the amorphous phase. System snapshots in the (b) BCC phase, HCP phase with (c)straight or (d) rippled planes and in (e) the amorphous solid phase. (f) RDF along the isobar $P^* = 0.28$ for different temperatures in the amorphous phase (cyan lines) and in the HDL phase (blue lines).}
    \label{solidphases}
\end{figure}

In a ($P,\;\rho$) phase diagram, we can estimate the critical point location using the isothermal density derivatives of the pressure \cite{corradini2010}

\begin{equation}
\left (\frac{\partial P}{\partial \rho}  \right)_{T}=\left (\frac{\partial^2 P}{\partial \rho^2}  \right )_{T}=0\;.
\end{equation} 

Coming from the supercritical region, the LLCP lays at the end of the Widom Line (WL) - a line in the $(P, T)$ phase diagram that may be acquired by the extension of the LLPT curve into the one-phase supercritical region (separation between LDL-like and HDL-like behavior) and is the locus of maximum fluctuations of the order parameter \cite{holten2012, bianco2019}. In this work we are estimating the WL based on the $\kappa_T$ maxima. In the figure~\ref{waterPTdiagram} (a) we show the pure CSW potential phase diagram obtained from our simulations. The WL, indicated by the dotted purple line and the purple squares, ends at the LLCP. Below the LLCP we have the transition between the liquid phases, indicated by the discontinuity in the thermodynamic property $\kappa_T^*$ -- shown in the ESI$^\dag$ for all isotherms -- and in the structural and dynamic properties. For instance, the upper panel in the figure~\ref{waterPTdiagram}(b) shows the translational order parameter $\tau^*$ for the subcritical pressure $P^* = 0.13$ and for the supercritical pressure $P^* = 0.15$ as function of the inverse of temperature. As $T^*$ decreases we may see a discontinuity in the subcritical isobar, indicating an abrupt change in the fluid structure. On the other hand, the supercritical isobar has a monotonic increase in $\tau^*$ as $T^*$ decreases, indicating an increase in the particles order. Similarly, the dependence of the diffusion coefficient $D^*$ with $T^{*-1}$ is discontinuous in the supercritical isobar. For the supercritical pressure $P^* = 0.15$ we see a change in the diffusion slope with temperature when it crosses the Widom Line, indicating the HDL-dominated to LDL-dominated regime. These results are in agreement with our recent work obtained by the heating of the system~\cite{marques2020} and previous works employing this potential~\cite{franzese2007,alan2008,franzeseJCP2010,franzese2011}.

However, the solid phases weren't quite explored for this system. In fact, the hexagonal closed packed (HCP) phase was observed in our work~\cite{marques2020} and in the study by Hus and Urbic using the methanol model~\cite{urbic2013, urbicPRE2014b}; in confined environments, Leoni found that the CSW potential forms different layered crystals that resemble the HCP planes\cite{leoni2014}. Here, exploring a larger region in the phase diagram, we characterized three distinct solid phases. The solid phase I corresponds to a body-centered cubic (BCC) crystal at lower pressures. Increasing $P^*$ it changes to the solid phase II, with a HCP structure, and at even higher compression we observe the amorphous solid, named phase III. The transition between the solid phases, and from LDL to HCP, are well defined by the discontinuous behavior in $\kappa_T^*$, shown in the ESI. The transition from solid phase I to LDL and from solid phase II to HDL have discontinuities in the response functions $\alpha_p^*$ and $C_p^*$, shown in the ESI. Also, the structure (here characterized by the pair excess entropy) and the dynamic behavior (given by the diffusion coefficient) are discontinuous for these solid-fluid transition. This can be observed in the figure~\ref{waterPTdiagram}(c). Here, the pressure $P^* = 0.05$ is a isobar that cross the BCC-LDL transition, $P^* = 0.11$, 0.14 and 0.20 the HCP-HDL transition and $P^* = 0.28$ the amorphous-HDL boundary. The structure and dynamics change smoothly for this border, as the magenta line for $P^* =0.28$ indicates in the figure~\ref{waterPTdiagram} (c). Also, the response functions $C_p^*$ and $\alpha_p^*$ are not discontinuous in this limit, but have a maximum. This is indicative of the amorphous-HDL boundary, which may change if the system is going through a cooling or a heating process, as the amorphous phase is not an equilibrium one. The distinct phases can also be observed when we analyze the water-water radial distribution function (RDF) $g_{ww}(r^*)$ along one isotherm. For instance, we show in the figure~\ref{solidphases}(a) the $g_{ww}(r^*)$ for pressures ranging from $P^* = 0.01$ to $P^* = 0.30$ along the isotherm $T^* = 0.26$. We can see clear changes in the structure as $P^*$ varies. At lower pressures the particles are separated mainly at the second length scale, characterizing the BCC phase -- a snapshot at $P^* = 0.01$ and $T^* = 0.26$ is shown in the figure~\ref{solidphases}(b). Increasing the pressure the system changes for the HCP phase, where the occupation in the first length scale dominates the structure. In fact, the HCP planes are separated by a distance equal to the second length scale, while the distance between particles in the same plane is the first length scale. It becomes clear when we use the Ovito~\cite{ovito} feature "create bonds" that the distance is equal to the first scale. While for the BCC snapshot we did not see any bond, for the HCP snapshot at $P^* = 0.14$ in the figure~\ref{solidphases}(c) we may see the bonds between particles in the same plane. As $P^*$ grows, the HCP planes get rippled, as we can see in figure~\ref{solidphases}(d) for $P^* = 0.24$ and in the behavior of the RDF red lines, in resemblance to the stripes wiggling found by Leoni~\cite{leoni2014}. Here, it is important to address that the distortion observed in the wrinkled HCP phase may be a consequence of our cubic cell: a strain is imposed over the crystalline structure if it did not match the cubic symmetry imposed by the cubic cell with periodic boundary conditions. This artificial strain acts as a confinement, leading to the deformation observed in the HCP structure - and it was noticed in the confined case~\cite{leoni2014}. Finally, it changes to an amorphous structure at high pressure, as we show for $P^* = 0.30$ in figure~\ref{solidphases}(e). In the figure 3(f) it is possible to see how the $g_{ww}(r^*)$ behaves along the isobar $P^* = 0.28$. At low temperatures, the system is in the amorphous phase, with lower mobility in comparison to the HDL phase - figure 2(c). Increasing the temperature, the system crosses the amorphous-HDL boundary and structural changes can also be observed: the first peak becomes small and the second peak changes its shape. Also, the amorphous phase indicates longer range peaks, while the HDL phase does not. Even the break between the first two peaks indicates a change in system structure: there is no occupancy between the first two peaks for the amorphous, similarly to a solid, unlike the HDL curves, which show occupancy in the region between the first two peaks - as in a fluid.

Now, with the phase behavior of the pure CSW potential model depicted, we may perceive how the presence of short alcohol affects the observed phases and the water-like anomalies.

\subsection{Water-short alcohol mixtures}
\begin{figure}[h!]
    \centering
    \subfigure[]{\includegraphics[width=0.475\textwidth]{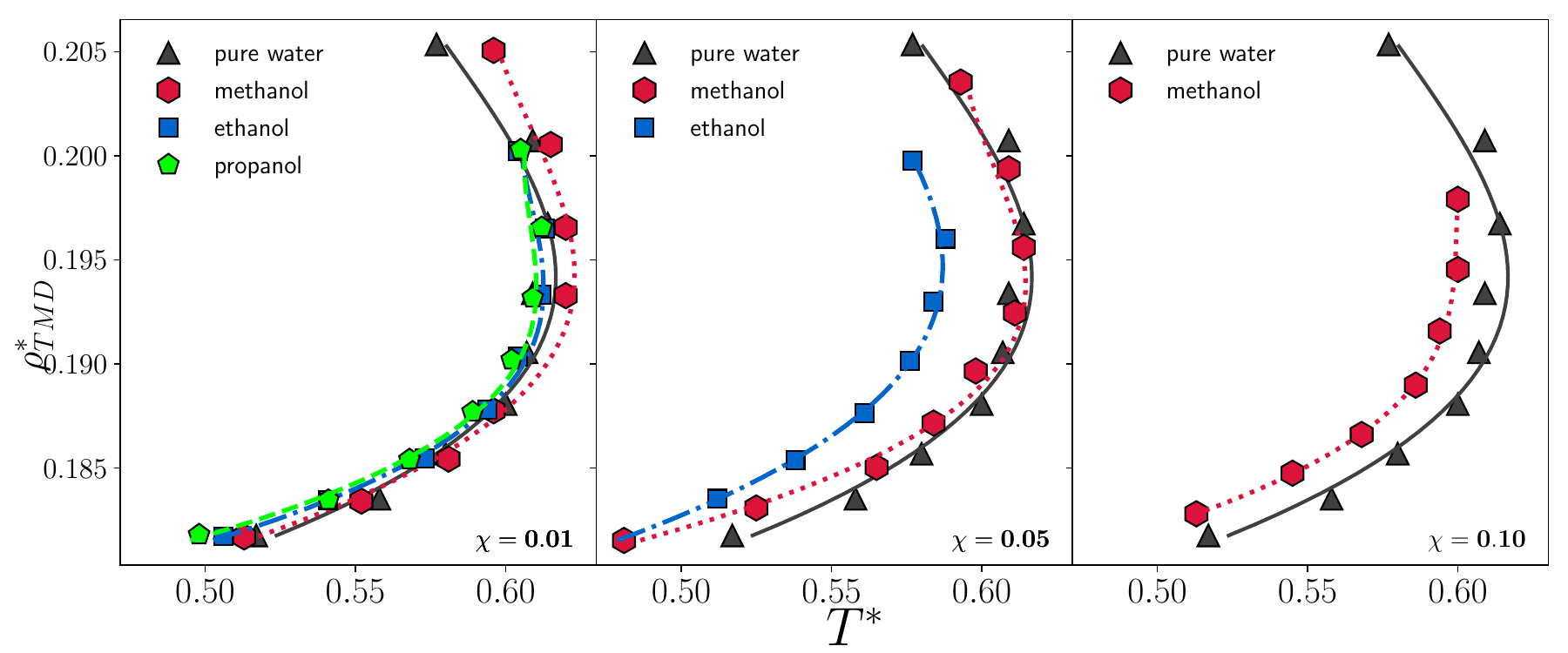}}
    \centering
     \subfigure[]{\includegraphics[width=0.1675\textwidth]{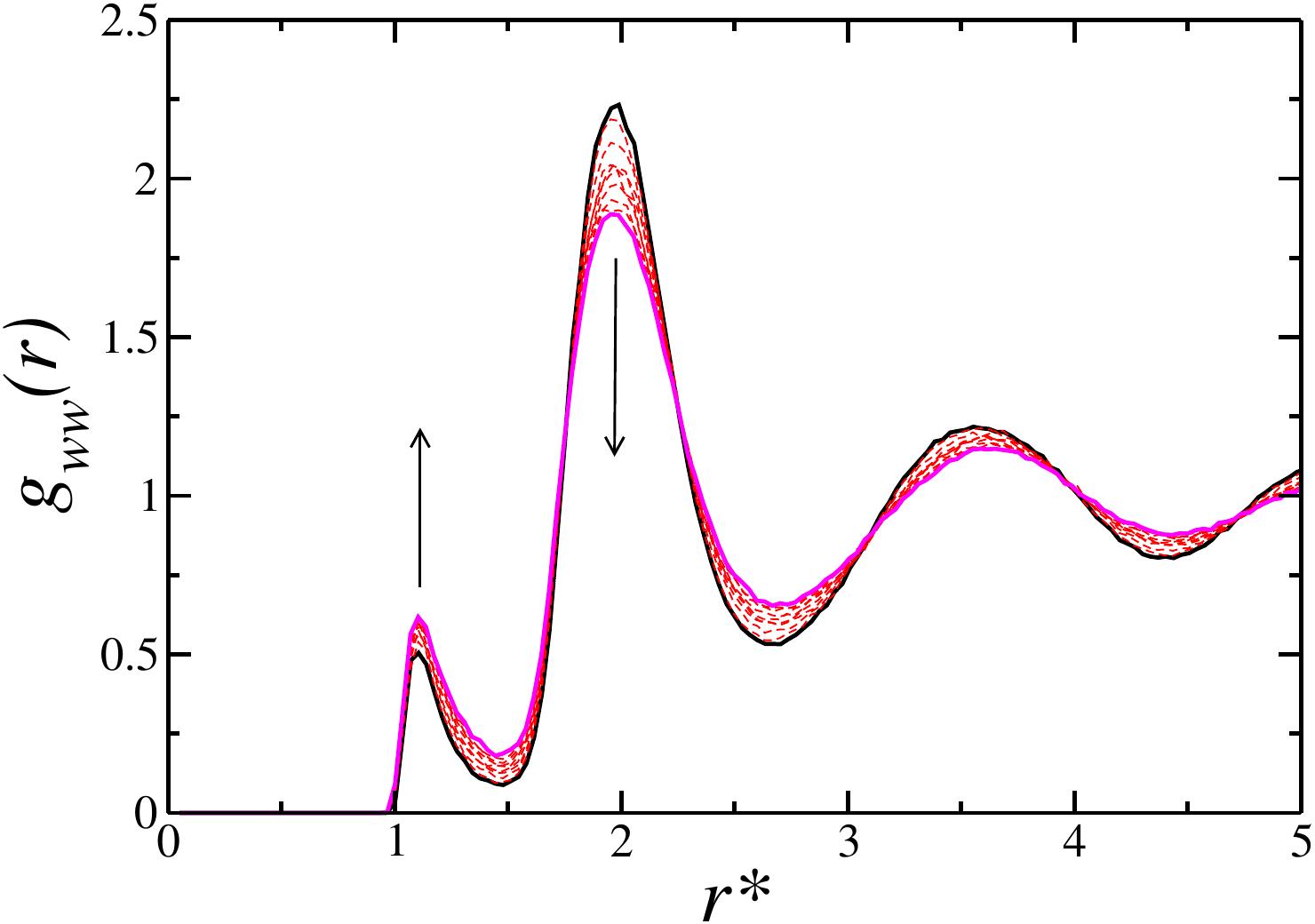}}
    \centering
     \subfigure[]{\includegraphics[width=0.1675\textwidth]{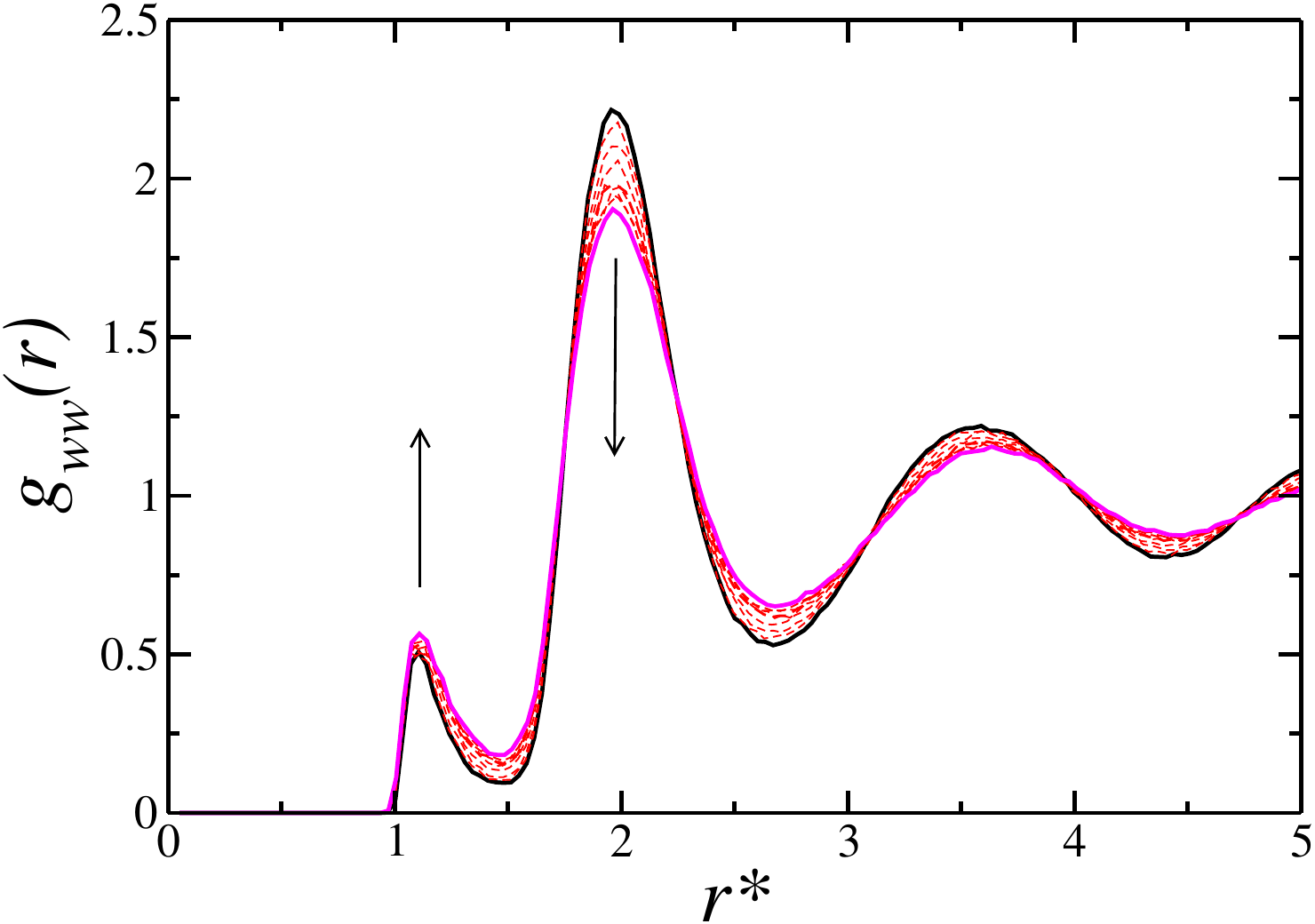}}
     \centering
     \subfigure[]{\includegraphics[width=0.1675\textwidth]{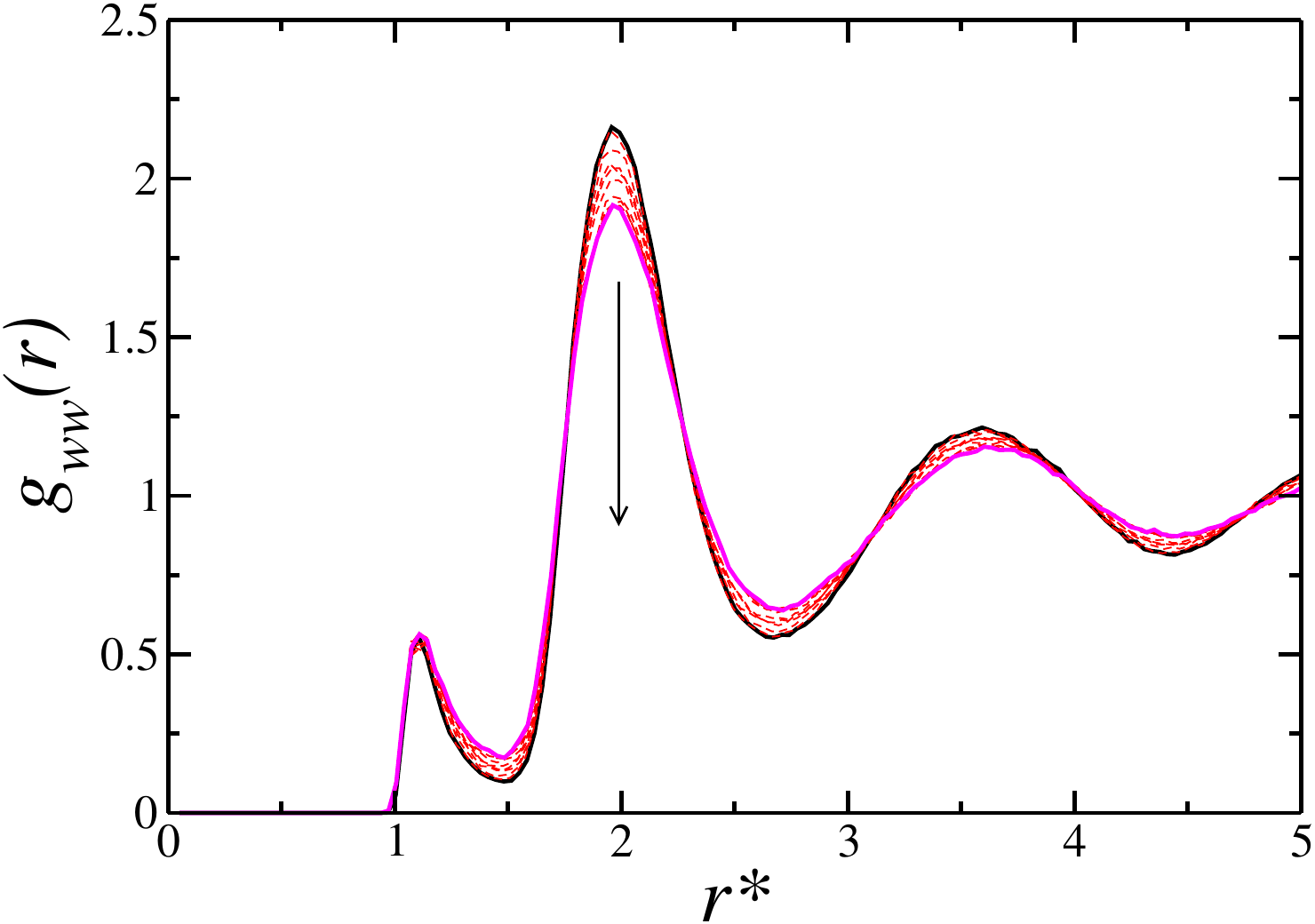}}
     \caption{(a) TMD behaviour of all CS alcohols used in this work: for $\chi_{alc}=0.05$, 1-propanol didn't show TMD and for $\chi_{alc}=0.10$, only methanol shows TMD. (b) CSW-CSW radial distribution function $g_{ww}(r^*)$ along the isobar $P^* = 0.08$ for water-ethanol at $\chi_{alc}=0.01$ case with temperatures ranging from $T = 0.50$ (black solid line) to  $T^* = 0.68$ (magenta solid line). The intermediate temperatures are shown with red dashed lines. The arrows indicate the competition between the scales as $T$ increases. (c) is for the case with ethanol concentration at $\chi_{alc} = 0.05$, were the competition between the scales and the TMD are still observable, while for (d) $\chi_{alc} = 0.10$ both competition and TMD vanish.}
    \label{tmd_allalcohols}
\end{figure}

Small clusterings of short chain alcohols such as methanol, ethanol and 1-propanol create a very interesting effect in the TMD line~\cite{wada1962a,Salgado20}. They act as "structure maker", promoting the low density ice-like water structure and increasing the TMD. This is usually observed for alcohol concentrations $\chi_{alc}$ smaller than 0.01 - and is not our goal here. We want to analyze the TMD vanishing and what happens in the phase diagram as it vanishes. This can be observed as $\chi_{alc}$ increases and methanol acts as "structure breaker". Using the CS model for water-methanol mixtures~\cite{marques2020} we found that the TMD persists up to high methanol concentrations, as $\chi < 70$\% -- much higher than in experiments. This is a consequence of the model: the same potential is employed for water-water, water-OH and OH-OH. On the other hand, the energy for the water-water h-bonds is equal to the water-OH h-bonds. Nevertheless, we can increase the structure breaker effect by increasing the solute size. In fact, for ethanol the TMD line vanishes at $\chi_{alc} = 0.10$, while for 1-propanol the TMD is only observed at $\chi_{alc} = 0.01$ - the TMD lines are shown in the figure~\ref{tmd_allalcohols}(a). 

\begin{figure}[h!]
    \centering
     \subfigure[]{\includegraphics[width=0.2325\textwidth]{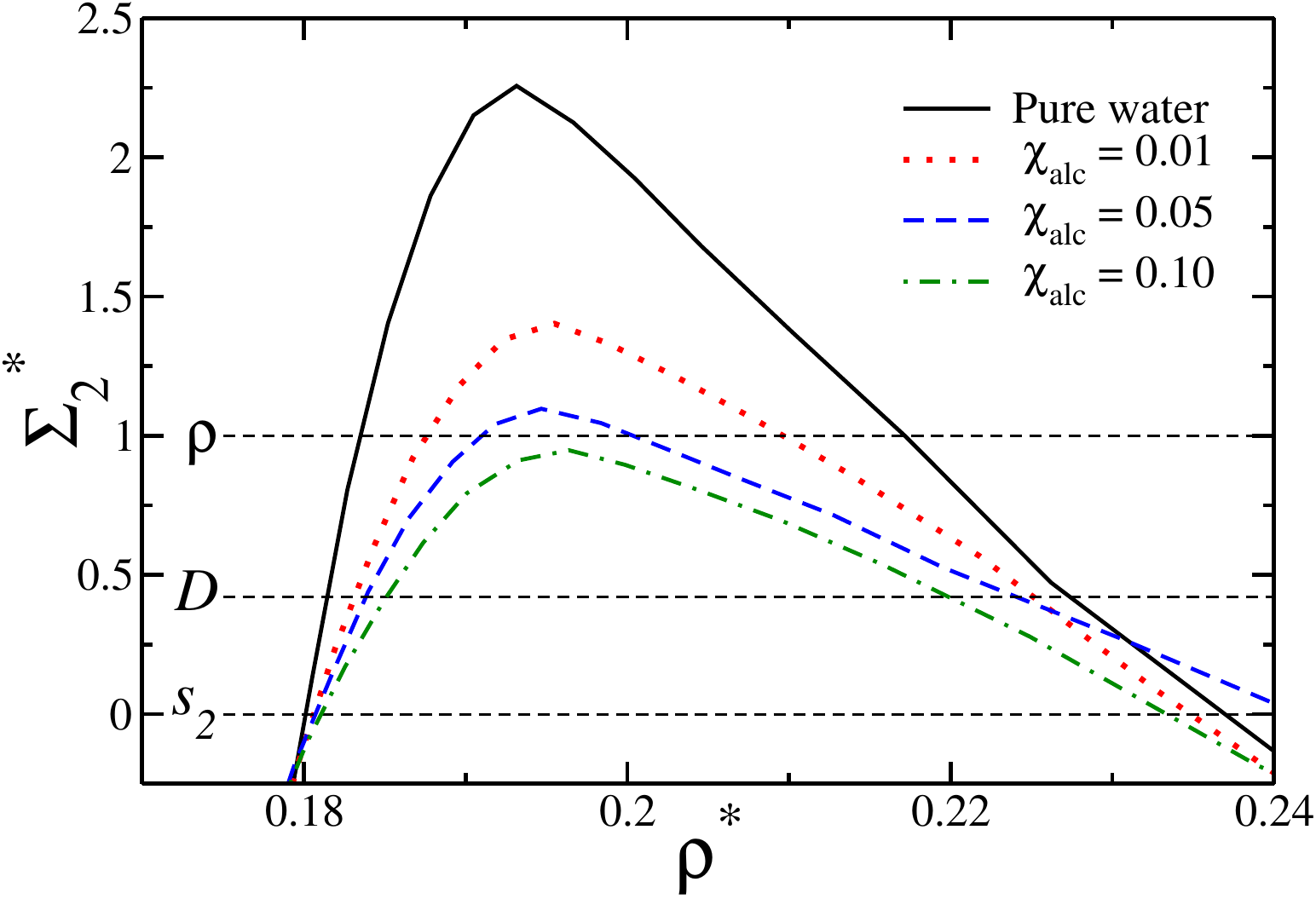}}
     \subfigure[]{\includegraphics[width=0.225\textwidth]{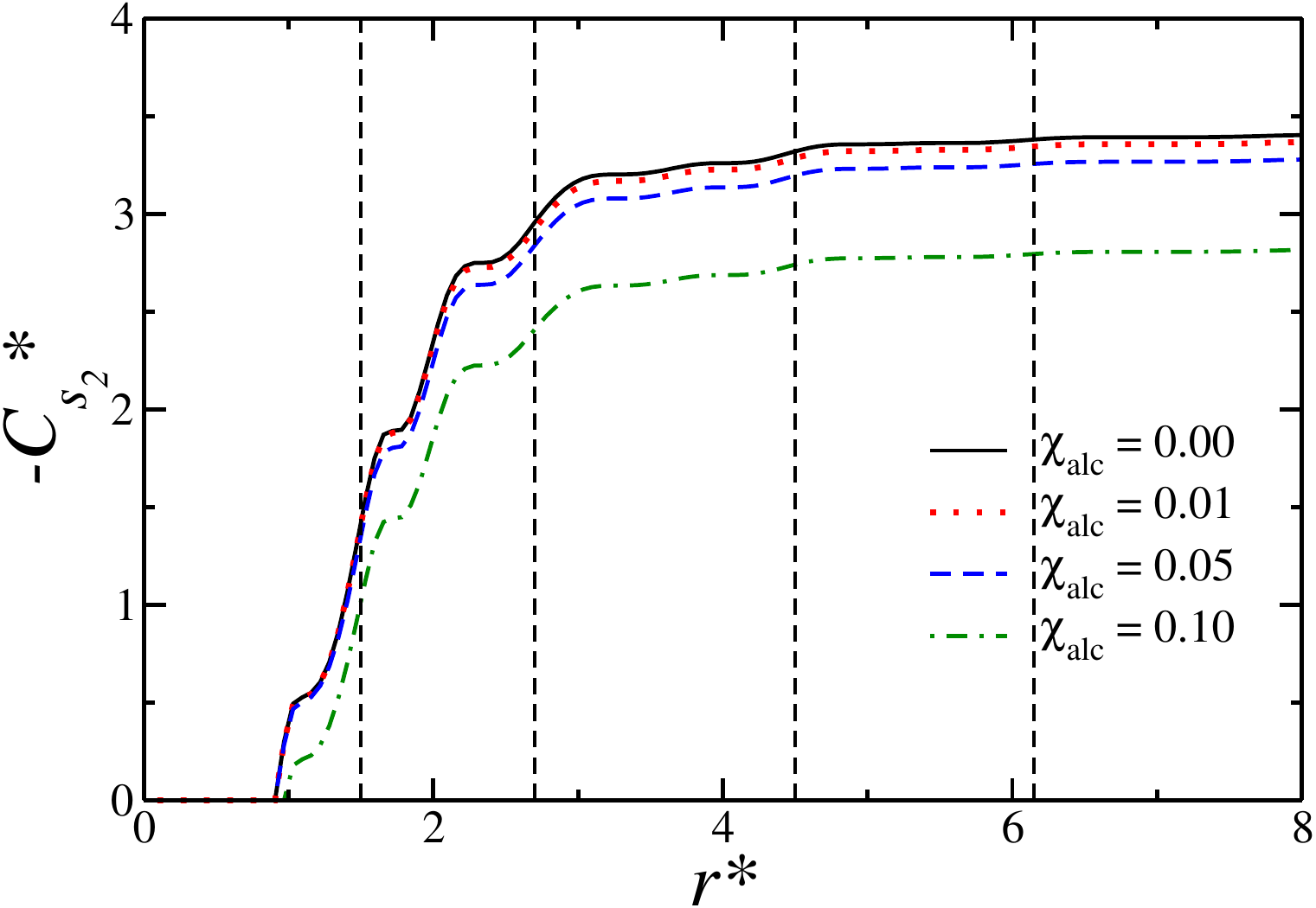}}
     \subfigure[]{\includegraphics[width=0.4\textwidth]{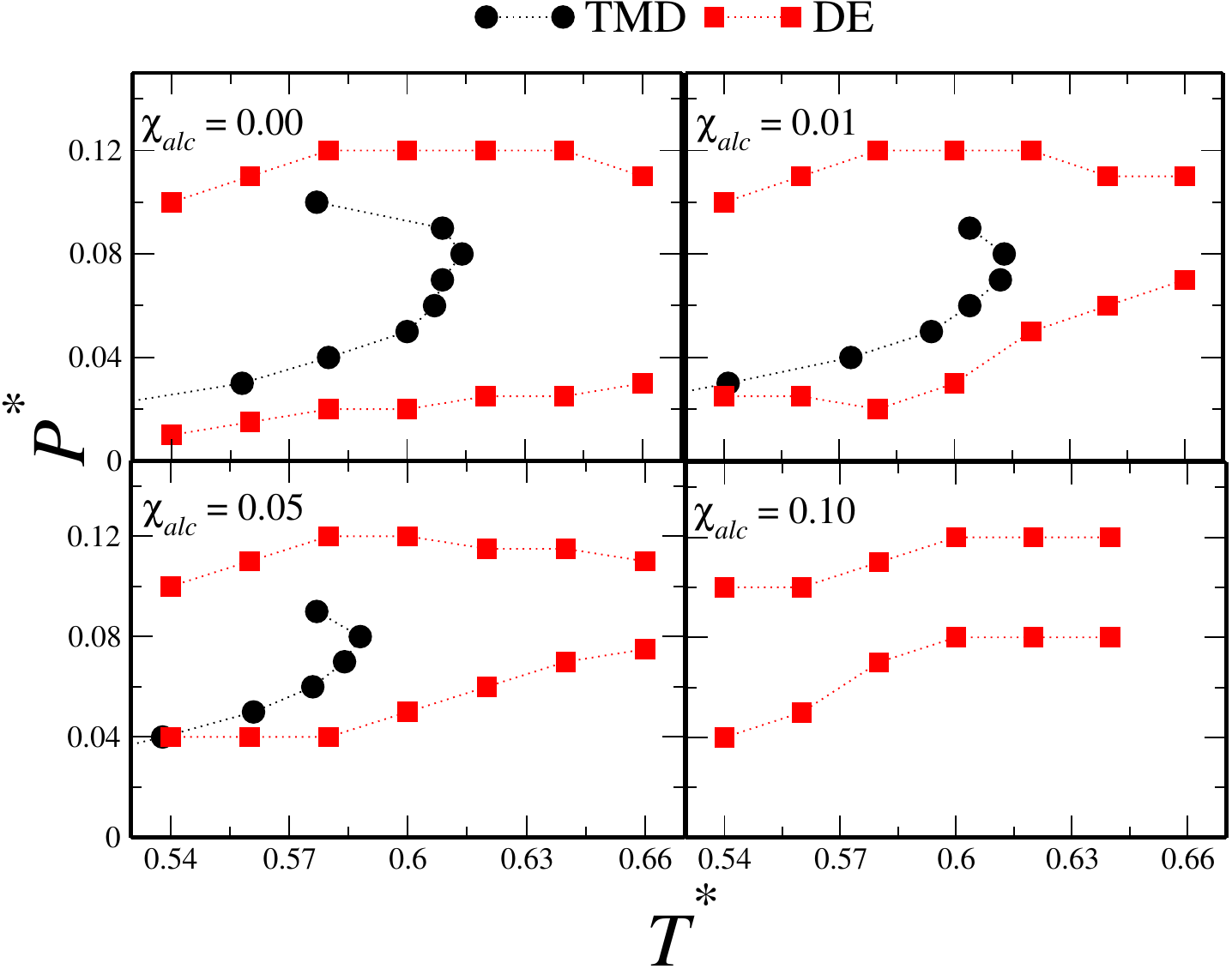}}
    \caption{(a) $\Sigma_2^*$ for water-ethanol mixtures at $T^* = 0.58$ showing the regions for the density ($\rho^*$), diffusion ($D^*$) and structural ($s_2^*$) anomalies. (b) Cumulative two-body entropy for the system at $T^* = 0.58$ and $P^* = 0.10$. The grey lines corresponds to the minima in the $g_{ww}(r^*)$ and delimit the coordination shells. (c) Diffusion extrema (DE) and TMD lines for distinct CSW ethanol concentrations, indicating that the anomalies gradually vanish as $\chi_{alc}$ increases.}
    \label{sigma_cum}
\end{figure}

The water anomalous behavior is related to the competition between two liquids that coexist~\cite{gallo2016,Evy13,marques2020a}. This competition can be observed using the $g_{ww}(r^*)$. Here we show the RDFs for the isobar $P^* = 0.08$ between the temperatures $T^* = 0.50$ and $T^* = 0.68$ for a fraction of ethanol of $\chi_{alc} = 0.01$ in the figure~\ref{tmd_allalcohols}(b), for $\chi_{alc} = 0.05$ in the figure~\ref{tmd_allalcohols}(c) and for $\chi_{alc} = 0.10$ in figure~\ref{tmd_allalcohols}(d). For the fractions $\chi_{alc} = 0.01$ and $\chi_{alc} = 0.05$ we observe the competition between the scales: the water particles migrates from the second length scale to the first length scale as $T^*$ increases, as indicated by the arrows. On the other hand, for $\chi_{alc} = 0.10$ there is practically no increase in the occupation of the first length scale as the occupation in the second length scale decreases. Since there is no competition, we do not observe the density anomaly. This indicates that adding higher concentration of alcohol changes the competition between the scales in the CSW water model and that the CS alcohol chain length also affects the competition - in our previous work for methanol, we only observe this at $\chi_{alc} = 0.70$~\cite{marques2020}. In contrast, the anomalies can occur due to competitions in the long range coordination shells - not only in the first or second one~\cite{Krek08,franzeseJCP2010} . Such behavior may be scrutinized by means of $\Sigma_2$, defined as~\cite{sharma2006, alan2008}

\begin{equation}
\Sigma_2 = \left(\frac{\partial s_2}{\partial \ln\rho}  \right)_T\;.
\end{equation}

\begin{figure}[h!]
\centering
    \includegraphics[width=0.48\textwidth]{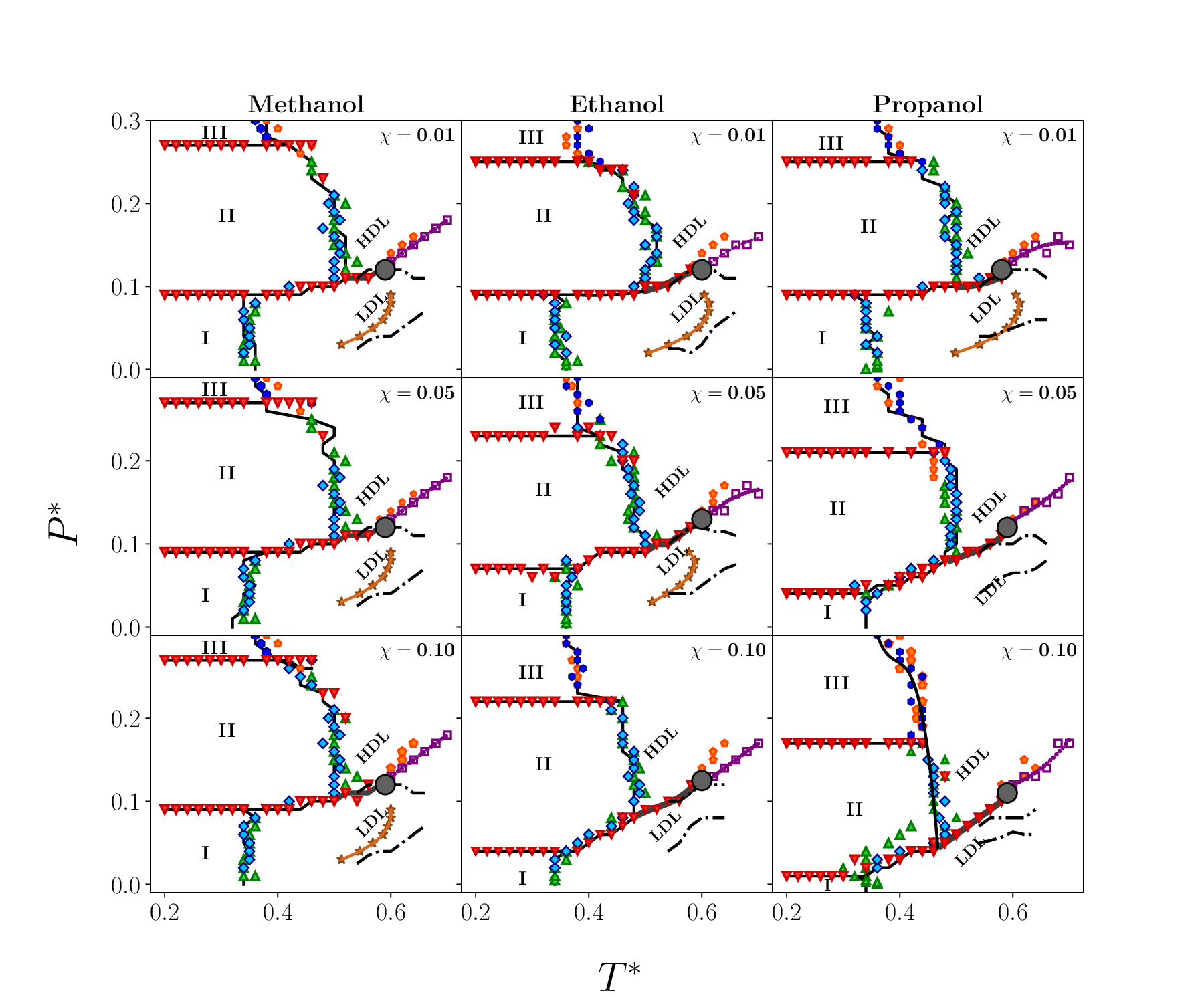} 
    \caption{$PT$ phase diagrams for aqueous solutions of (left) methanol, (center) ethanol and (right) propanol for all concentrations analyzed in this work.}
    \label{PTdiagrams}
\end{figure}

\begin{figure}[ht!]
    \centering
    \includegraphics[width=0.485\textwidth, height=0.425\textwidth]{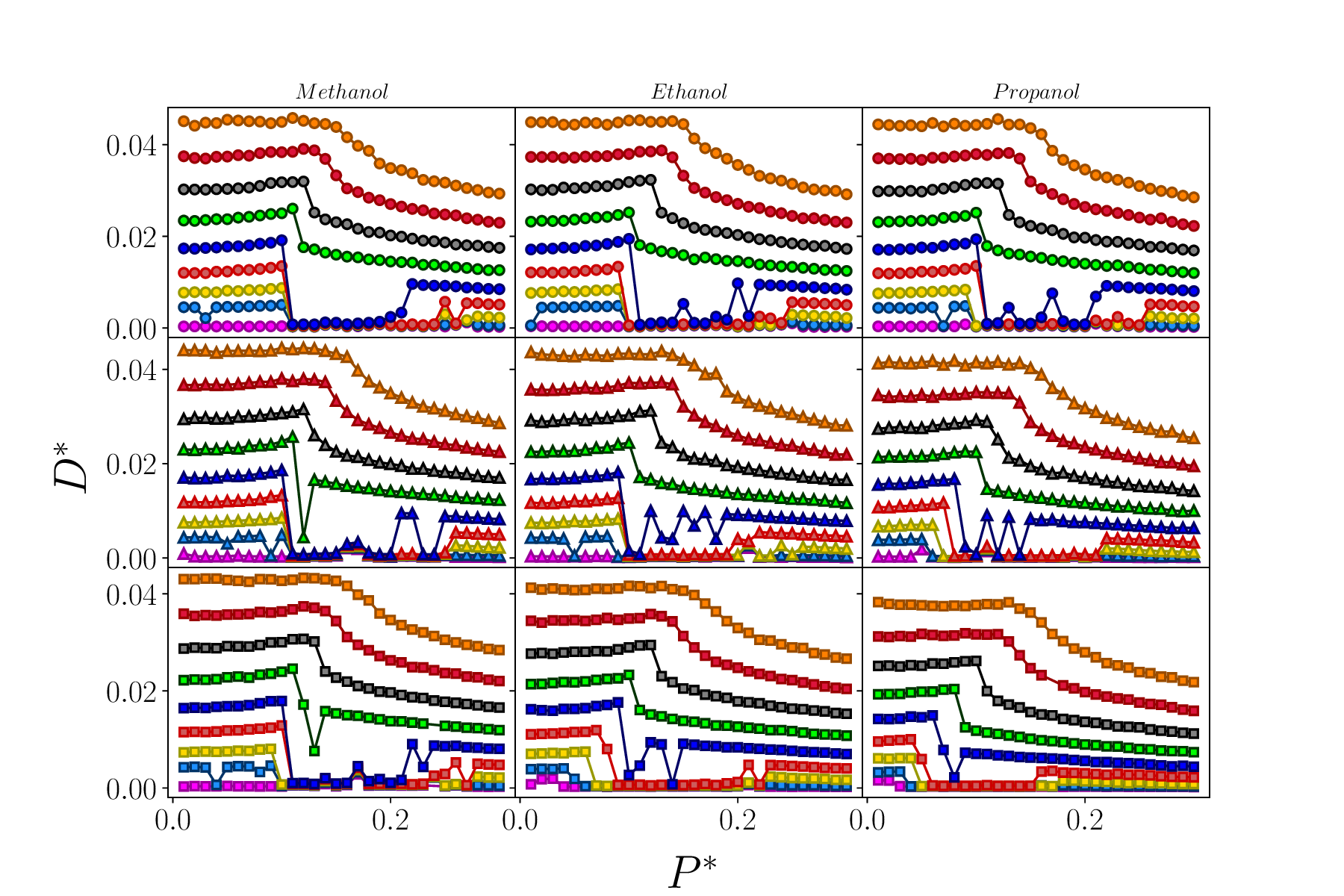}
    \caption{Diffusion coefficient versus pressure for all solutions analyzed in this work. From bottom to top in each diagram, we have the isotherms $T^*=0.34,0.38,0.42,...0.66$. Each row represents a concentration of solute: $\chi_{alc}=0.01,0.05\;and\;0.10$}
    \label{Dallfracs}
\end{figure}

As Errington and co-authors have shown~\cite{Er06}, and others authors corroborated for the CSW model~\cite{alan2008, franzese2010}, $\Sigma_2^* > 1.0$ is the thermodynamic condition for existence of the density anomaly. In this way, in the figure~\ref{sigma_cum}(a) we show the $(\Sigma_2^* ,\rho^*)$ along the isotherm $T^* = 0.58$ for the pure CSW system and mixtures with ethanol molar fractions $\chi_{alc} = 0.01$, $0.05$ and $0.10$. Adding solute, the thermodynamic condition isn't satisfied and the TMD gradually vanishes. However, the thermodynamic condition for other anomalies are still satisfied. A system will have diffusion anomaly if $\Sigma_2^* > 0.42$. As indicated in figure~\ref{sigma_cum}(a) all cases have dynamics anomaly -- the diffusion extrema are indicated in the phase diagrams of figure~\ref{PTdiagrams} and the $(D^*,P^*)$ isotherms are exhibited in the figure~\ref{Dallfracs}. However, we observed that the diffusion anomaly region shrinks as longer are the alcohols chains and higher are the concentrations. This indicates a gradual disappearance of the anomalies, an effect that has already been investigated by Vilaseca and Franzese~\cite{franzese2010}: by tuning the potential softness, they have noticed a predominance of short-range over long-range structures, leading to the gradual extinction of the anomalies. In order to ascertain  how the short and long range CSW coordination shells are affected by the solute, we evaluated the cumulative two-body entropy~\cite{Krek08,Cardoso21} 
\begin{equation}
\label{cs2}
C_{s_2}(r) = -2\pi\rho \int_0^r [g(r')\ln(g(r')) - g(r') +1]r'^2 dr'\;.
\end{equation}
Here, $r$ is the upper integration limit. Because the two-body excess entropy is a structural order metric that connects thermodynamics and structure, $C_{s_2}$ is a useful tool to analyze structural characteristics of the core-softened system. This is presented for distinct ethanol concentrations in the figure~\ref{sigma_cum}(b) at $T^* = 0.58$ and $P^* = 0.10$. The dashed lines indicate the minima in the $g_{ww}(r^*)$. Therefore, they divide the coordination shells. Looking for the lower alcohol concentration case, $\chi_{alc} = 0.01$, there is no significant difference from the pure water case, $\chi_{alc} = 0.00$, in the first two coordination shells, although  it's possible to see a decrease in the order in the longer-range shells. This can be related to  decreasing in the TMD and DE region, figure~\ref{sigma_cum}(c). Increasing $\chi_{alc}$ to 0.05 leads to a disorder in the second shell, which now has a $C_{s_2}$ smaller than the pure-CSW case. As a consequence, the TMD and DE regions reduce their extension. Finally, the TMD disappearance is observed when even the first coordination shell structure is affected, as it can be seen for $\chi_{alc} = 0.10$. This illustrates a gradual disappearance of the anomalous behavior when the solute concentration or the non-polar chain size increases: the solute initially have an effect on the long-range coordination shells, decreasing the TMD region in phase diagrams until its complete disappearance points out that the short-range ordering is favored. Whereas the short-range ordering is favored by addition of more alcohol molecules to or the growth of the non-polar alcohol chain in the system, there is less competition between short- and long-range ordering and the anomalies start to vanish. The same gradual disappearing of the anomalies was previously reported by Vilaseca and Franzese~\cite{franzese2010}, where they changed the parameter $\Delta$ in the CSW interaction, Eq. (\ref{franzese}), favoring the short-range ordering and, as consequence of the lack of competition, killing the anomalies. This indicates that increasing the amount of alcohol makes the competition uneven - reinforcing that without competition between two fluids there are no anomalies.

In this work, all cases have LLPT with a LLCP, revealing that a system can have a LLCP without have density anomaly. This result is known since 2001 when the squared version of the CSW potential was studied in Franzese et al. \cite{Franzesenature2001}. The mechanism ruling the absence of the density anomaly in the CSW model was studied in Ref. ~\cite{franzese2010}, where it was shown that, however, other anomalies were present. After the liquid-liquid critical point, The Widom Line (WL) separates water with more HDL-like local structures at high temperatures from water with more LDL-like local structures at low temperatures \cite{malamace2009}. Looking at the diffusion coefficient $D^*$ isotherms, distinct transitions are observed. At lower temperatures as $T^* = 0.34$, it melts from the solid phase to the HDL phase at high pressures (figure~\ref{Dallfracs}). Increasing $T^*$, we may sight the system going from the LDL phase (with $D^* > 0$) to the HCP phase ($D^* \sim 0$), and to the HDL phase, were $D^*$ increases again. In the isotherms that cross the LDL-HDL transition, the discontinuity foreshadows a phase transition. Above the LLCP we may notice a change in the $(D^*, P^*)$ curve behavior as it crosses the WL. 

\begin{figure}[ht!]
    \centering
    \includegraphics[width=0.475\textwidth]{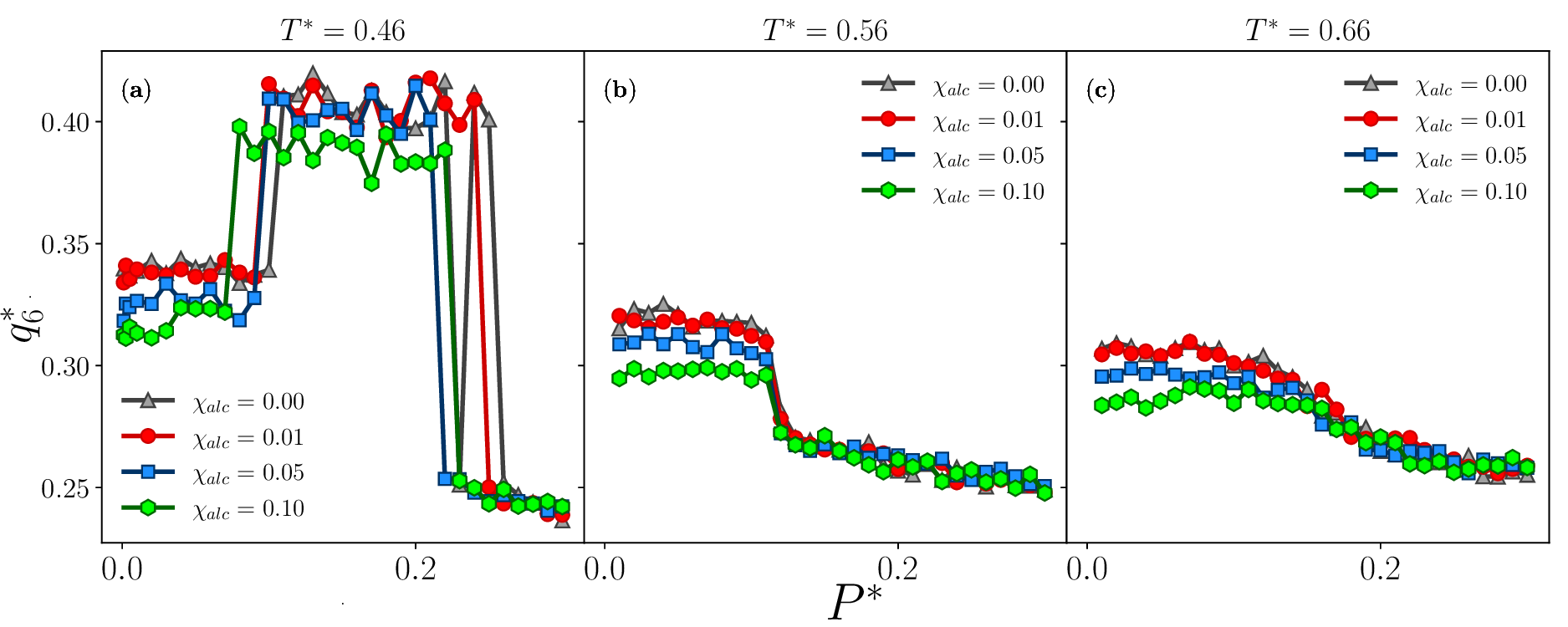}
     \caption{Local bond orientation order parameter $q_6^*$ as function of pressure for the water-ethanol isotherms (a) $T^* = 0.46$, (b) $T^* = 0.56$ and (c) $T^* = 0.66$.}
    \label{q6_llpt}
\end{figure}

The same discontinuity can be observed in the structural behavior. Besides $\tau$ and $s_2$, we also evaluated the $q_6$ to analyze the structure. For simplicity, we show here only the case of pure water and water-ethanol mixtures. Along the isotherm $T^* = 0.46$, that crosses the phases LDL, HCP and HDL, we may recognize in the figure~\ref{q6_llpt}(a) clearly the discontinuities for these transitions for all ethanol molar fractions. It is also noticeable how the ethanol is structure breaker: the $q_6$ values in the LDL and HCP phases decrease with the growth of the ethanol molar fraction. Along the isotherm $T^* = 0.56$, shown in the figure~\ref{q6_llpt}(b), the same structure breaker effect in the LDL regime and a discontinuous transition to the HDL phase are present. In contrast, above the LLCP there is no discontinuity. The isotherm $T^* = 0.66$, shown in the figure~\ref{q6_llpt}(c), has a smooth decay in $q_6$ when it crosses the WL and goes from LDL-like to HDL-like. 

\begin{figure}[ht]
        \centering
     \subfigure[]{\includegraphics[width=0.15\textwidth]{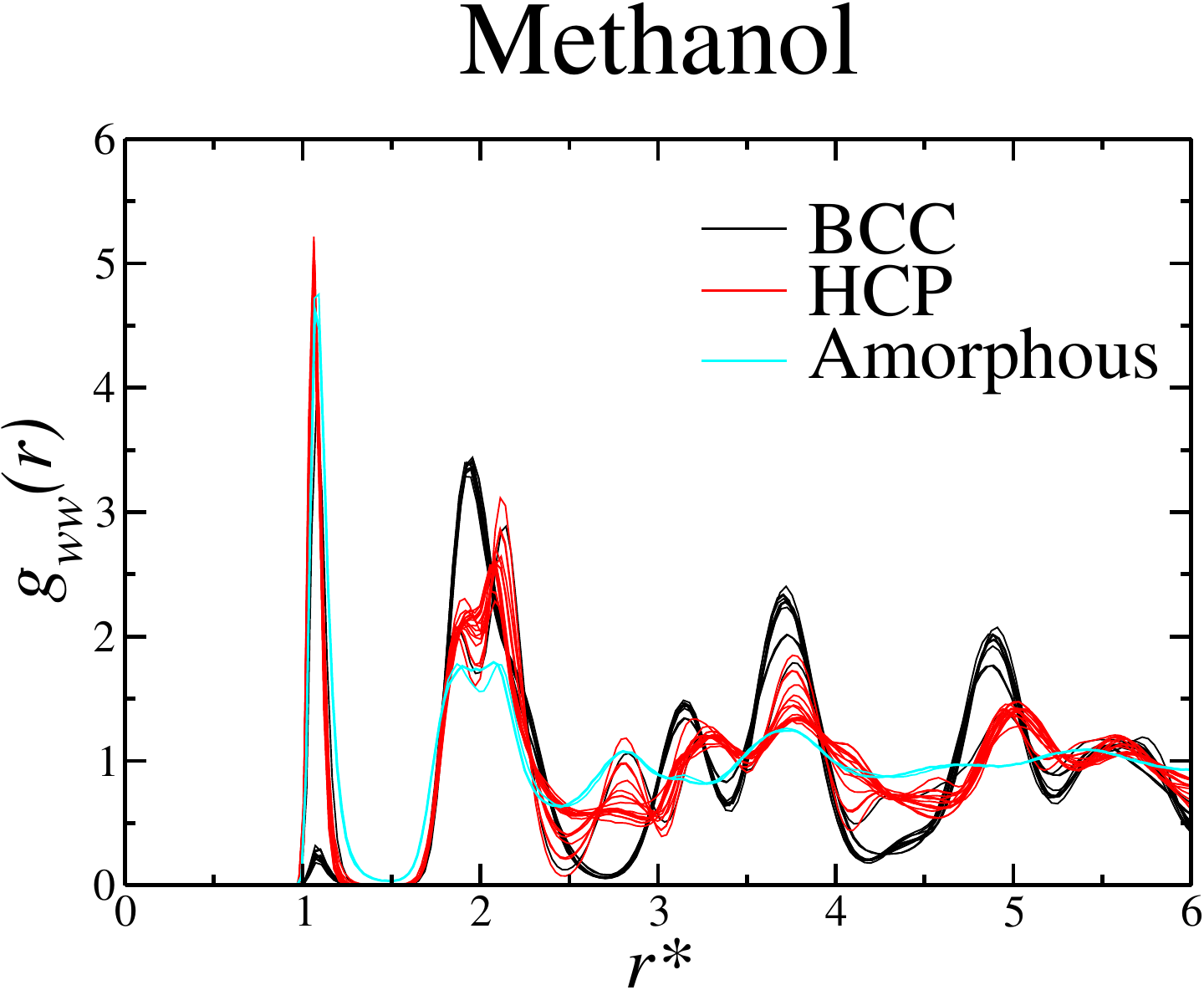}}
    \centering
     \subfigure[]{\includegraphics[width=0.15\textwidth]{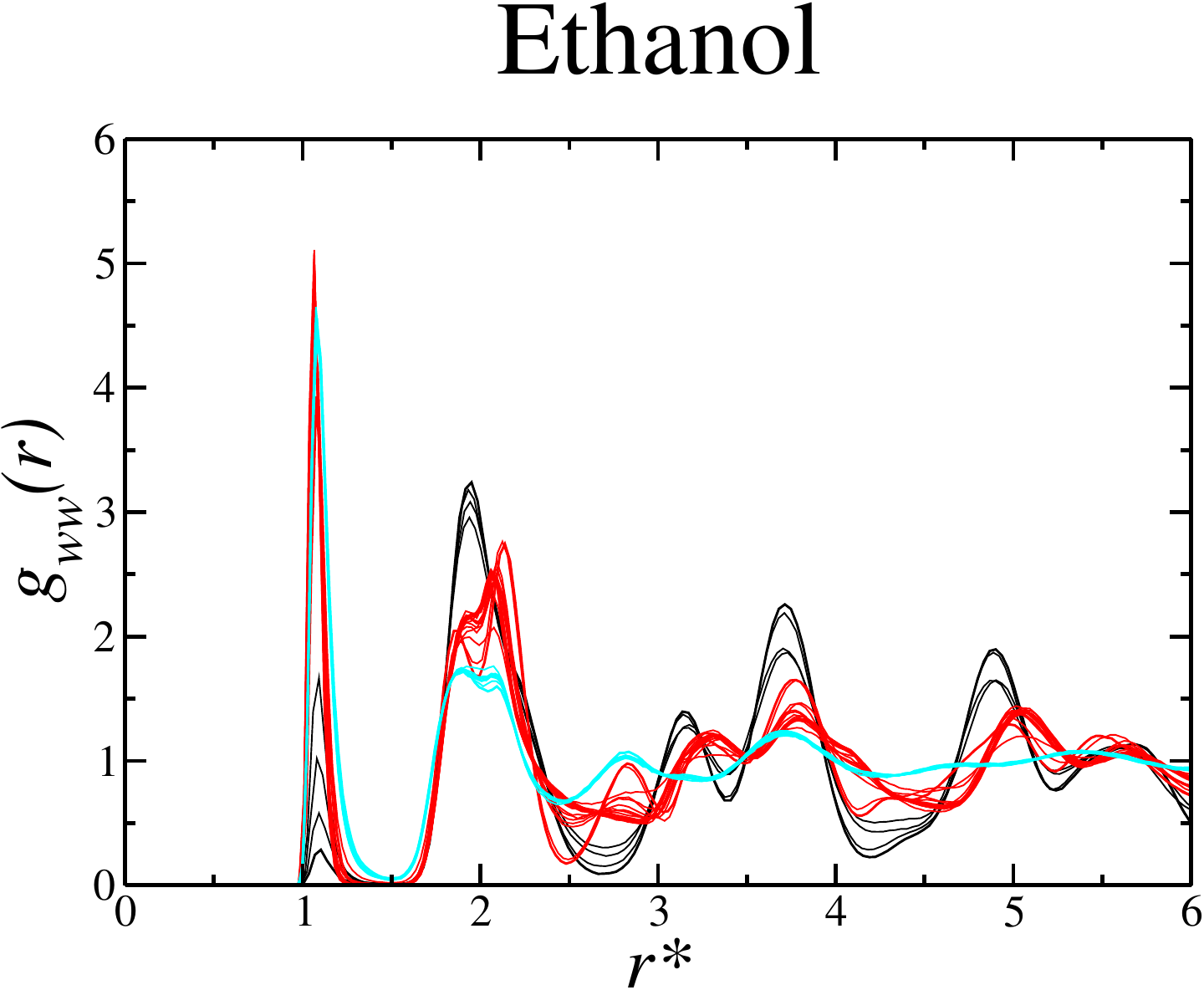}}
     \centering
     \subfigure[]{\includegraphics[width=0.15\textwidth]{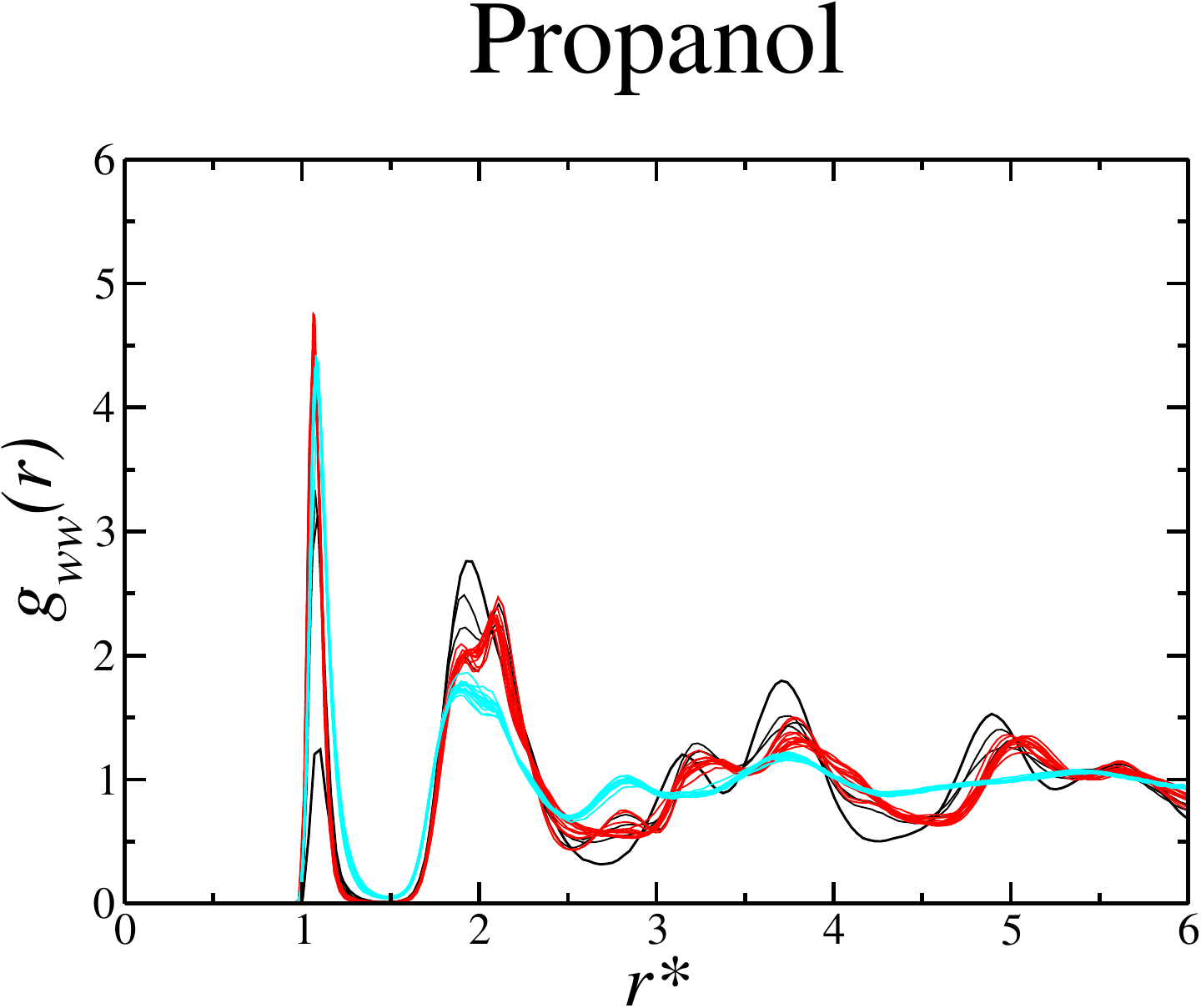}}
     \centering
     \subfigure[]{\includegraphics[width=0.14\textwidth]{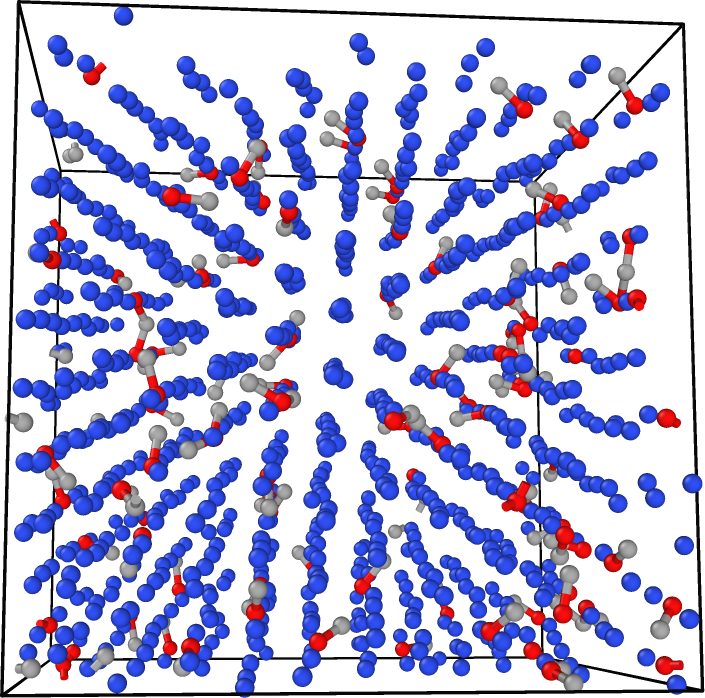}}
     \centering
     \subfigure[]{\includegraphics[width=0.14\textwidth]{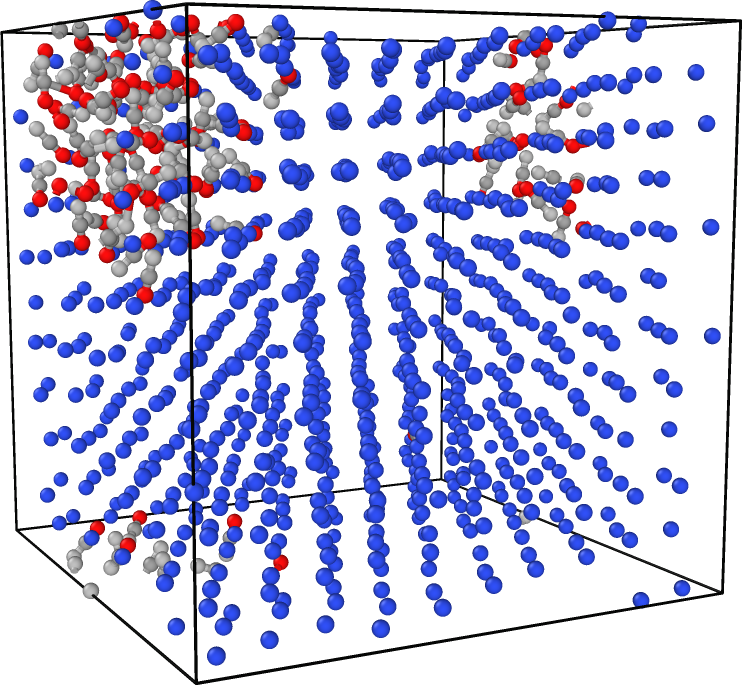}}
     \centering
     \subfigure[]{\includegraphics[width=0.14\textwidth]{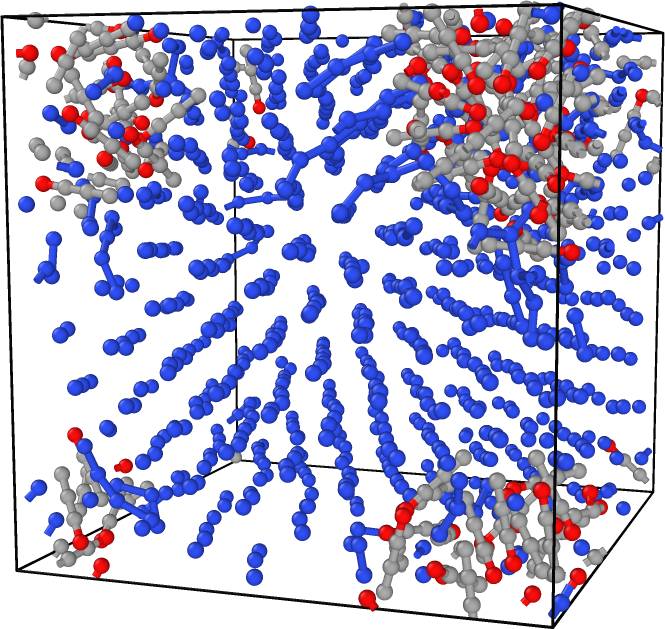}}
     \centering
     \subfigure[]{\includegraphics[width=0.3\textwidth]{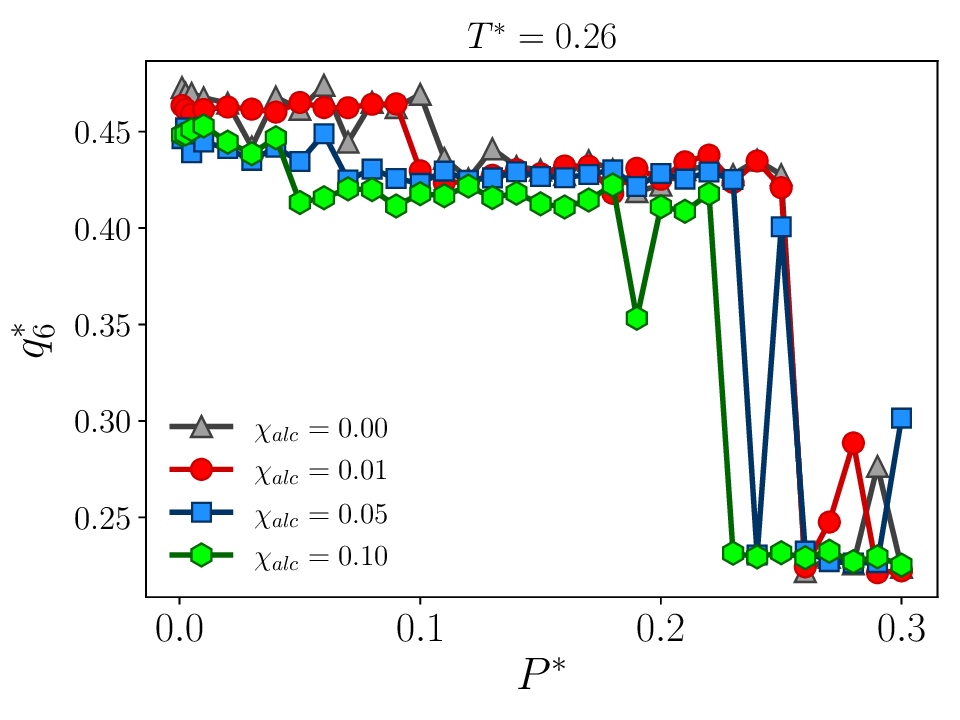}}
    \caption{ CSW-CSW radial distribution function $g_{ww}(r^*)$ along the isotherm $T^* = 0.26$ for water-alcohol mixtures with $\chi_{alc}=0.10$ for (a) methanol, (b) ethanol and (c) propanol. Black lines stands for the BCC phase, red lines for the HCP solid and cyan lines for the amorphous phase. Snapshots for the correspondents mixtures: $\chi_{alc}=0.10$ of (d) methanol, (e) ethanol and (f) propanol. (g) Local bond orientation order parameter $q_6$ as function of pressure for the isotherm $T^* = 0.26$ and distinct concentrations of ethanol.}
    \label{rdf_solid_alcohol}
\end{figure}

As in the pure CSW potential case, three solid phases were observed in the phase diagrams representing the mixtures. A compelling finding is how the carbon chain length and the solute concentration affect the solid phases, modifying their relative arrangement: the BCC crystal (region I) loses space and moves to lower pressures; the HCP crystal (region II) also tends to shift towards lower pressures, without changing its total area occupied in the $PT$ phase diagrams. For these two solid phases, the temperature range in the $PT$ phase diagram seems to be independent of the fraction $\chi_{alc}$. On the other hand, the amorphous solid phase (region III) is favored since it expands its extension to lower pressures and higher temperatures as the non-polar chain grows. To understand why the HCP phase remains occupying a large area in the phase diagrams while the BCC area shrinks, we show in the figure~\ref{rdf_solid_alcohol} the $g_{ww}(r^*)$ for the three alcohols with fraction $\chi_{alc}=0.10$ along the isotherm $T^* = 0.26$ from $P^* = 0.01$ to $P^* = 0.30$. Comparing the three cases with the pure CSW case, figure~\ref{solidphases}(a), is clear that increasing the size of the alcohol non-polar tail favors the occupation in the second length scale. While for methanol and ethanol we see a low occupation in the first length scale at the lower pressures, for  1-propanol this occupancy is high even for $P^* = 0.01$. This is consequence of an alcohol bubble formation: for the CS methanol, since the molecule size is comparable to the second length scale and the OH is modeled as a CSW particle, the molecules merge in the BCC structure, as we show in the figure~\ref{rdf_solid_alcohol}(d). However, the longer alcohols create bubbles, as explained in the figure~\ref{rdf_solid_alcohol}(e) and (f) for ethanol and propanol, respectively. They modify the water structure in their vicinity, favoring the first length scale and reflecting in the local orientation. In the figure~\ref{rdf_solid_alcohol}(g) we show the $q_6$ for this isotherm for the case of pure CSW potential ($\chi_{alc} = 0.00$) and the three fractions of ethanol. For both crystal phases, BCC and HCP, the local order is affected by the alcohol because $q_6$ is small for higher fractions.

\section{Conclusions}
In this paper we have explored the supercooled regime of pure water and mixtures of water and short chain alcohols: methanol, ethanol and propanol using a two-length scale core-softened potential approach. Our aim was to understand the influence of chain size on anomalies, the liquid-liquid phase transition and the polymorphism which are generally observed in these models. There's a pronounced influence of the non-polar chain size on solid polymorphism. The BCC phase decreases its extension in the phase diagrams since longer alcohols are supporting the occupancy in the first length scale. As the HCP phase has also a higher occupancy in this first length scale, it is shifted to smaller pressures, while the amorphous solid phase grows by being favored by alcohol-induced disorder. 

The density anomaly vanishes as the competition between the scales is suppressed in the LDL phase. However, the LDL-HDL phase transition and the diffusion anomaly persist for all cases. In the case of the diffusion anomaly, larger sizes of the non-polar chain or higher alcohol concentrations cause a reduction in its extension in the phase diagram. This indicates that the competition between two liquids is connected with waterlike anomalies, but the system will no necessarily have the density anomaly if it has a first-order liquid-liquid phase transition ending at a liquid-liquid critical point. To this (still) subtle relationship between the second critical point and the presence of waterlike anomalies, we add the solute size and concentration shift, which leads to the gradual disappearing of these anomalies through the progressive influence on the CSW fluid coordination shells, favoring the short-range ordering and uneven the competition. These results help to understand the complex behavior of water and mixtures with amphiphilic solutes in the supercooled regime.

\section*{Acknowledgements}
Without the public funding this research would be impossible. MSM thanks the Brazilian Agencies Conselho Nacional de Desenvolvimento Cient\'ifico e Tecnol\'ogico (CNPq) for the PhD Scholarship and Coordena\c c\~ao de Aperfei\c coamento de Pessoal de N\'ivel Superior (CAPES) for the support to the collaborative period in the Instituto de Química Fisica Rocasolano in Madrid, and Universidade Federal do Oeste da Bahia (UFOB) for the period of official leave for the doctorate. VFH thanks the CAPES, Finance Code 001, for the MSc Scholarship. JRB acknowledge the Brazilian agencies CNPq and Funda\c c\~ao de Apoio a Pesquisa do Rio Grande do Sul (FAPERGS) for financial support. All simulations were performed in the SATOLEP Cluster from the Group of Theory and Simulation in Complex Systems from UFPel.

\section{References}
\bibliography{solute_size_arxiv}
\end{document}